\documentclass{article}

\newcommand{\acro}[1]{\textsc{#1}\xspace}

\newcommand{\rrl}{\acro{rl}}
\newcommand{\grl}{\acro{grl}}
\newcommand{\acd}{\acro{acd}}
\newcommand{\gacd}{\acro{gacd}}
\newcommand{\iit}{\acro{it}}

\newcommand{\mdp}{\acro{mdp}}
\newcommand{\pomdp}{\acro{pomdp}}
\newcommand{\ppo}{\acro{ppo}}
\newcommand{\ot}{\acro{ot}}
\newcommand{\vgae}{\acro{vgae}}
\newcommand{\mlp}{\acro{mlp}}

\newcommand{\cage}{\acro{cc2}}

\newcommand{\psg}{\acro{psg}}
\newcommand{\gtm}{\acro{gtm}} 
\newcommand{\gnn}{\acro{gnn}}
\newcommand{\mpnn}{\acro{mpnn}}

\newcommand{\fgw}{\acro{fgw}}
\newcommand{\mdh}{\acro{mdh}}
\newcommand{\mse}{\acro{mse}}
\newcommand{\aeot}{\acro{ae}-\ot}
\newcommand{\sdot}{\acro{sdot}}
\newcommand{\sota}{\acro{sota}}
\newcommand{\gt}{\acro{gt}}

\newcommand{\E}{\mathbf{E}}
\newcommand{\V}{\mathbf{V}}

\newcommand{\G}{\mathbf{G}}
\newcommand{\X}{\mathbf{X}}
\newcommand{\Z}{\mathbf{Z}}
\newcommand{\z}{\mathbf{z}}
\newcommand{\x}{\mathbf{x}}
\newcommand{\tmap}{\mathcal{T}}

\newcommand{\loss}{\mathcal{L}}
\newcommand{\graphs}{\{ G_t \}_{t \in [T]}}
\newcommand{\argmin}{\operatornamewithlimits{argmin}}

\usepackage{microtype}
\usepackage{graphicx}
\usepackage{epstopdf}
\usepackage{subcaption}
\usepackage{booktabs} 

\usepackage{hyperref}

\usepackage[accepted]{icml2025}

\usepackage{amsmath}
\usepackage{amssymb}
\usepackage{mathtools}
\usepackage{amsthm}
\usepackage{placeins}

\usepackage{diagbox}

\usepackage[capitalize,noabbrev]{cleveref}
\crefformat{section}{\S#2#1#3}
\crefformat{subsection}{\S#2#1#3}
\crefformat{subsubsection}{\S#2#1#3}

\usepackage{markdown} 
\usepackage{enumitem}
\usepackage{xspace}
\usepackage{arydshln}
\usepackage{footmisc}
\usepackage{tikz}
\usepackage{float}
\usepackage{pgf,pgfplots,pgfplotstable}
\pgfplotsset{compat=1.18}
\usetikzlibrary{shapes,decorations,arrows,calc,arrows.meta,fit,positioning,trees,backgrounds,patterns,bending}
\usepackage{xifthen}
\tikzset{
    node distance =1 cm and 1 cm,
    semithick,
    state/.style ={ellipse, draw, minimum width = 0.7 cm},
    point/.style = {circle, draw, inner sep=0.04cm,fill,node contents={}},
    bidirected/.style={Latex-Latex,dashed},
    el/.style = {inner sep=2pt, align=left, sloped}
}
\newcommand{\circled}[1]{\textcircled{\raisebox{-0.9pt}{#1}}}

\usetikzlibrary{shadows}
\tikzset{%
  cascaded/.style = {%
    general shadow = {%
      shadow scale = 1,
      shadow xshift = 2ex,
      shadow yshift = 2ex,
      draw,
      fill = white},
    general shadow = {%
      shadow scale = 1,
      shadow xshift = 1ex,
      shadow yshift = 1ex,
      draw,
      fill = white},
    fill = white,
    draw,
    minimum width = 4cm,
    minimum height = 2cm}
}

\theoremstyle{plain}
\newtheorem{theorem}{Theorem}[section]

\theoremstyle{definition}
\newtheorem{definition}[theorem]{Definition}

\theoremstyle{remark}

\DeclareMathOperator*{\argmax}{argmax} 

\usepackage[textsize=tiny]{todonotes}

\icmltitlerunning{General Autonomous Cybersecurity Defense}

\begin{document}

\twocolumn[
\icmltitle{General Autonomous Cybersecurity Defense:\\Learning Robust Policies for Dynamic Topologies and Diverse Attackers}



\icmlsetsymbol{equal}{*}

\begin{icmlauthorlist}
\icmlauthor{Arun Ramamurthy}{equal,yyy}
\icmlauthor{Neil Dhir}{equal,xxx,cran}
\end{icmlauthorlist}

\icmlaffiliation{yyy}{SIEMENS, Princeton, NJ, USA.}
\icmlaffiliation{xxx}{Work performed while ND was an employee of SIEMENS.}
\icmlaffiliation{cran}{Cranfield University, Bedford, UK}

\icmlcorrespondingauthor{Arun Ramamurthy}{arun.ramamurthy@siemens.com}
\icmlcorrespondingauthor{Neil Dhir}{neil.dhir@cranfield.ac.uk}

\icmlkeywords{Cybersecurity, reinforcement learning, graph neural networks, optimal transport, multi-task learning, network security}

\vskip 0.3in
]



\printAffiliationsAndNotice{\icmlEqualContribution} 

\newcommand\note[1]{\textcolor{red}{#1}}

\begin{abstract}
    In the face of evolving cyber threats such as malware, ransomware and phishing, autonomous cybersecurity defense (\acd) systems have become essential for real-time threat detection and response with optional human intervention. However, existing \acd systems rely on limiting assumptions, particularly the stationarity of the underlying network dynamics. In real-world scenarios, network topologies can change due to actions taken by attackers or defenders, system failures, or time evolution of networks, leading to failures in the adaptive capabilities of current defense agents. Moreover, many agents are trained on static environments, resulting in overfitting to specific topologies, which hampers their ability to generalize to out-of-distribution network topologies. This work addresses these challenges by exploring methods for developing agents to learn generalizable policies across dynamic network environments -- general \acd{} (\gacd).
\end{abstract}
\section{Introduction}
\label{sec:intro}


\begin{figure}[ht!]
    \centering
    \resizebox{\columnwidth}{!}{
        \input{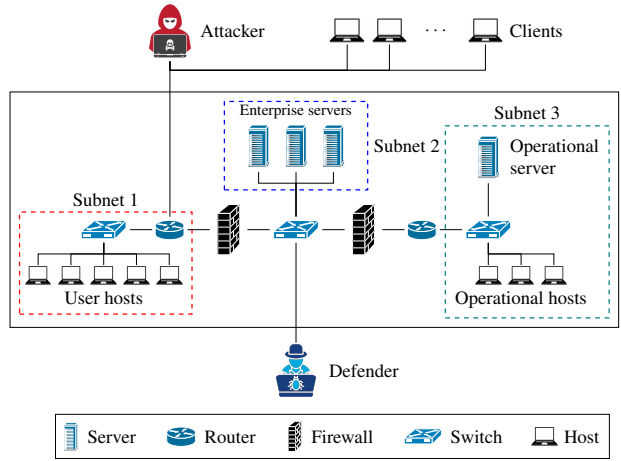}
    }
    \caption{\iit network of the \textsc{cage} challenge two (\cage) scenario  \citep{cage_challenge_2}. An organization's \iit infrastructure and the actors involved in the use case. The operator, the \emph{defender}, of this infrastructure takes measures to protect the infrastructure against an \emph{attacker} while providing services to a client population.}
    \label{fig:standard_defender}
\end{figure}

In today's digital landscape, cybersecurity is paramount for individuals, organizations and governments. With the proliferation of sophisticated cyber threats such as malware, ransomware and phishing attacks, there is a pressing need for robust and active defense mechanisms to safeguard sensitive data and systems. \acd systems have emerged as a critical component in the ongoing battle against cyber threats, offering real-time detection, analysis and response capabilities with optional human intervention \citep{dhir2021prospective}. A growing trend has been to treat the interaction between the attacker (red agent), the defender (blue agent) and the environment (which includes green agents i.e. regular users) as a reinforcement learning problem \citep{hammar2020finding, ridley2018machine}.

The field of \acd is vast and growing, but existing work faces several limiting assumptions. First, it is common to assume stationarity in the underlying system dynamics. In most real-world scenarios this assumption is easily violated. For example, an attacker's, defender's or user's action often results in changes to the network; infrastructure evolves due to functional upgrades and innovation \citep{hammar2020finding}, a mechanical failure in the system, etc. -- alter the underlying network topology. This in turn causes a distribution shift of the observations available to \acd agents. Blue agents who rely on vectorized representations of the observation spaces fail to adapt to the non-stationary environment.

Similarly, in an orthogonal setting, current network defense agents are trained and deployed to optimally defend a \emph{single} scenario defined by a static network topology, as shown in \cref{fig:standard_defender}, which represents the \iit network of the \textsc{cage} challenge two (\cage) scenario \citep{cage_challenge_2} -- a popular cybersecurity challenge. As a result, defenders are overfitted to perform optimally in these scenarios, to the extent that even a simple reordering of the nodes in the network causes the agent to underperform in the same scenario (see \cref{sec:randomization}). Thus, we can conclude that the agent does not learn any meaningful semantics of the scenario during its training and does not learn a policy that can be transferred to new or unobserved scenarios or network topologies.

To address these limitations, we propose a framework for \emph{general} \acd{} (\gacd), leveraging the representational power of graph neural networks (\gnn) \citep{kipf2016variational,jiang2018graph} and the flexibility of optimal transport \citep{lei2019geometric, an2019ae, an2020ae, chen2020graph}. 

\paragraph{Contributions}
\begin{itemize}
    \item We develop a \gacd agent that learns to generalize its policy across various network topologies by integrating graph embeddings with proximal policy optimization (\ppo) \citep{schulman2017proximal}.
    \item We demonstrate the feasibility (and utility) of encoding any desired set of network topologies into a continuous space, devoid of discontinuities between topologies, using optimal transport.
    \item We train \gacd agents to generalize over numerous network topologies and adversaries, i.e. red agents, in a multi-task learning setting.
    \item We empirically show that our agent optimally learns to generalize its network defense by adaptively and minimally sampling the network topology space, ensuring efficient exploration and exploitation.
\end{itemize}


\section{Preliminaries}
\label{sec:problem_statement}


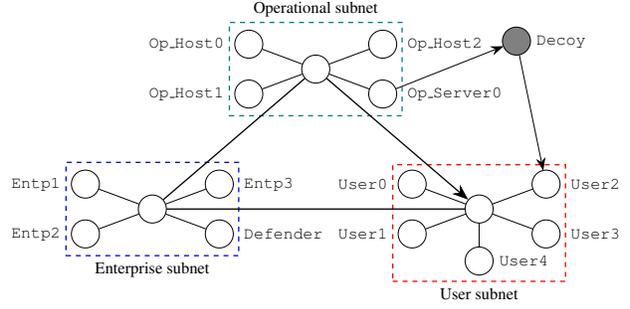
\begin{figure*}[ht!]
    \centering
    \begin{subfigure}[b]{0.49\textwidth}
        \centering
        \caption{Example state $s_t$, of the \cage environment.}
        \label{table:challenge_2_state}
        \resizebox{\columnwidth}{!}{
            \begin{tabular}{ccccc}
            \toprule
            \textbf{Subnet} & \textbf{IP Address} & \textbf{Hostname} & \textbf{Activity} & \textbf{Compromised} \\
            \midrule
                \texttt{10.0.17.48/28} & \texttt{10.0.17.55} & \texttt{Defender} & None & No \\
                \texttt{10.0.17.48/28} & \texttt{10.0.17.51} & \texttt{Enterprise0} & None & User \\
                \texttt{10.0.17.48/28} & \texttt{10.0.17.56} & \texttt{Enterprise1} & None & No \\
                \texttt{10.0.17.48/28} & \texttt{10.0.17.49} & \texttt{Enterprise2} & Exploit & User \\
            \midrule
                \texttt{10.0.78.16/28} & \texttt{10.0.78.19} & \texttt{Op\_Host0} & None & No \\
                \texttt{10.0.78.16/28} & \texttt{10.0.78.22} & \texttt{Op\_Host1} & None & No \\
                \texttt{10.0.78.16/28} & \texttt{10.0.78.21} & \texttt{Op\_Host2} & None & No \\
                \texttt{10.0.78.16/28} & \texttt{10.0.78.28} & \texttt{Op\_Server0} & None & No \\
            \midrule
                \texttt{10.0.93.128/28} & \texttt{10.0.93.134} & \texttt{User0} & None & No \\
                \texttt{10.0.93.128/28} & \texttt{10.0.93.137} & \texttt{User1} & None & No \\
                \texttt{10.0.93.128/28} & \texttt{10.0.93.132} & \texttt{User2} & None & No \\
                \texttt{10.0.93.128/28} & \texttt{10.0.93.133} & \texttt{User3} & None & No \\
                \texttt{10.0.93.128/28} & \texttt{10.0.93.136} & \texttt{User4} & None & User \\
            \bottomrule
            \end{tabular}
-        }
    \end{subfigure}
    \hfill
    \begin{subfigure}[c]{0.49\textwidth}
        \centering
        \resizebox{\columnwidth}{!}{
            \begin{tikzpicture}[
    node distance=10mm,
    my_node/.style={circle, draw, minimum size=6mm},
    arrow/.style={-{Stealth[scale=1.5]}}
]

\node [my_node, label={[label distance=0mm]above:{}}] at (0,0) (os) {};
\node [my_node,above left = 1mm and 10mm of os, label={[label distance=1mm]left:{\texttt{Op\_Host0}}}] (os1) {};
\node [my_node,below left = 1mm and 10mm of os, label={[label distance=1mm]left:{\texttt{Op\_Host1}}}] (os2) {};
\node [my_node,above right= 1mm and 10mm of os, label={[label distance=1mm]right:{\texttt{Op\_Host2}}}] (os3) {};
\node [my_node,below right= 1mm and 10mm of os, label={[label distance=1mm]right:{\texttt{Op\_Server0}}}] (os4) {};
\foreach \i in {os1, os2, os3, os4} {
    \draw (os) --  (\i);
}
\node[draw, dashed, teal, thick, minimum width=3.7cm, minimum height=2cm, label={[label distance=0mm]above:{Operational subnet}}] at (os) {};

\node [my_node, label={[label distance=0mm]above:{}}] at (-3.5,-3) (es) {};
\node [my_node,above left = 1mm and 10mm of es, label={[label distance=1mm]left:{\texttt{Entp1}}}] (es1) {};
\node [my_node,below left = 1mm and 10mm of es, label={[label distance=1mm]left:{\texttt{Entp2}}}] (es2) {};
\node [my_node,above right= 1mm and 10mm of es, label={[label distance=1mm]right:{\texttt{Entp3}}}] (es3) {};
\node [my_node,below right= 1mm and 10mm of es, label={[label distance=1mm]right:{\texttt{Defender}}}] (es4) {};
\foreach \i in {es1, es2, es3, es4} {
    \draw (es) --  (\i);
}
\node[draw, dashed, blue, thick, minimum width=3.7cm, minimum height=2cm, label={[label distance=0mm]below:{Enterprise subnet}}] at (es) {};

\node [my_node, label={[label distance=0mm]above:{}}] at (3.5,-3) (us) {};
\node [my_node,above left = 1mm and 10mm of us, label={[label distance=1mm]left:{\texttt{User0}}}] (us1) {};
\node [my_node,below left = 1mm and 10mm of us, label={[label distance=1mm]left:{\texttt{User1}}}] (us2) {};
\node [my_node,above right= 1mm and 10mm of us, label={[label distance=1mm]right:{\texttt{User2}}}] (us3) {};
\node [my_node,below right= 1mm and 10mm of us, label={[label distance=1mm]right:{\texttt{User3}}}] (us4) {};
\node [my_node,below = 5mm of us, label={[label distance=0mm]right:{\texttt{User4}}}] (us5) {};
\foreach \i in {us1, us2, us3, us4, us5} {
    \draw (us) --  (\i);
}
\node[draw, dashed, thick, red, minimum width=3.7cm, minimum height=2.5cm, label={[label distance=0mm]below:{User subnet}}] at (3.5,-3.3) {};

\draw[arrow, thick] (os) -- (us);
\draw[thick] (os) -- (es);
\draw[thick] (es) -- (us);

\node [my_node,fill=gray, label={[label distance=0mm]right:{\texttt{Decoy}}}] at (4.3,0.6) (dh) {};
\draw[arrow] (os4) -- (dh);
\draw[arrow] (dh) -- (us3); 

\end{tikzpicture}
        }
        \caption{Graph representation $G$ of the \cage network -- see reference state in \cref{table:challenge_2_state} -- as observed by the blue agent. Each node in the graph is attributed with 7 features, described in Sec. \ref{sec:embedding}, that is used as a representation of the state of the host and subnet by the model. Dashed colored boxes correspond to those shown in \cref{fig:standard_defender}.}
        \label{fig:cage_challenge_2_graph}
    \end{subfigure}
    \caption{Tabular and graphical representation of the \cage network environment. In \cref{fig:cage_challenge_2_graph} the graph representation $G$ illustrates the network topology as seen by the \gacd agent, where nodes represent devices, subnets and decoys, and edges denote network interfaces.
    In \cref{table:challenge_2_state} the corresponding state $s_t$ of the network is presented in tabular form, detailing the subnets, IP addresses, hostnames, activities and compromise status of each node in the environment.}
\end{figure*}

We treat the \acd{} scenario as a reinforcement learning (\rrl) problem where we model the interaction between the blue agent and an environment (e.g. \cref{fig:standard_defender}) as a Markov decision process (\mdp).
\begin{definition}[Markov decision process]
An \mdp is a tuple $M = \langle \mathcal{S}, \mathcal{A}, P, R, \alpha \rangle$ where: $\mathcal{S}$ is the set of states, representing all possible configurations of the environment; $\mathcal{A}$ is the set of actions available to the agent; $P: \mathcal{S} \times \mathcal{A} \times \mathcal{S} \to [0,1]$ is the transition probability function, where $P(s' \mid s, a)$ represents the probability of transitioning to state $s' \in \mathcal{S}$ from state $s \in \mathcal{S}$ after taking action $a \in \mathcal{A}$; $R: \mathcal{S} \times \mathcal{A} \to \mathbb{R}$ is the reward function, where $R(s, a)$ is the expected reward received after taking action $a \in \mathcal{A}$ in state $s \in \mathcal{S}$; $\alpha \in [0,1]$ is the discount factor, which determines the importance of future rewards relative to immediate rewards.
\end{definition}

In the \rrl setting, a blue agent interacts with an environment over a sequence of time steps $\{0, \ldots, T\} \triangleq [T]$. At each time step $t$, the agent observes the state of the environment $s_t$ and selects an action $a_t$ according to a policy $\pi_\theta(a_t \mid s_t)$, parameterized by $\theta$. The policy $\pi$ is a function that determines the appropriate action to take given the current state. Here, the state $s_t$ encapsulates all relevant environmental information at time $t$ and its distribution is governed by the network topology.

After executing the action $a_t$, the agent receives a reward $r_t = R(s_t, a_t)$ from the environment, which quantifies the quality or desirability of the state-action pair $(s_t, a_t)$. The agent’s objective is to learn a policy that maximizes the expected sum of discounted rewards over trajectories generated by the policy. A policy $\pi^*$ solves the \rrl \emph{problem} \citep{sutton2018reinforcement} by maximizing the expected cumulative reward in an \mdp over a finite horizon $T$:
\begin{equation}
    \pi^* = \argmax_{\pi} \mathbb{E}\left[\sum_{t=0}^T \alpha^t r_{t+1}\right].
\end{equation}
An \rrl \emph{algorithm} provides a method to compute or approximate $\pi^*$ \citep{hammar2020finding}. A widely used example of such an algorithm is \ppo.

An example of the environmental state at time $t$ is shown in \cref{table:challenge_2_state}. Therein the `activity' column gives us a high-level description (or indirect observation) of the red and green agents' actions, but the `activity' does not fully represent the explicit actions of all the interacting agents. Consequently, the true actions of the agents remain unobserved. As such we need to treat the problem as a partially observed \mdp (\pomdp) which is an extension of the \mdp \citep{aastrom1965optimal}. In the \pomdp setting the agent does not directly observe $s \in \mathcal{S}$ but instead the agent receives observations $o \in \mathcal{O}$ that provide partial or noisy information about the state. Formally, a \pomdp is a tuple $\langle \mathcal{S}, \mathcal{A}, P, R, \mathcal{O}, \Lambda, \alpha \rangle$ where $\Lambda: \mathcal{S} \times \mathcal{A} \times \mathcal{O} \to [0,1]$ is the observation probability function, where $\Lambda(o \mid s', a)$ represents the probability of observing $o \in \mathcal{O}$ given that the environment transitions to state $s' \in \mathcal{S}$ after taking action $a \in \mathcal{A}$.

\paragraph{Treating \iit infrastructure as a graph}An \iit infrastructure is naturally represented as a graph, where nodes correspond to network elements (e.g. servers, routers and end-points) and edges capture their connections. Attributes like firewall strength, bandwidth and security levels are encoded as node or edge features, enabling a structured and interpretable analysis. The directed nature of the graph aligns with real-world \iit systems, reflecting directional relationships such as data flow and access controls imposed by network administrators. Consequently, let $G = (\V,\E)$ denote a graph with a set of vertices $\V = \{V_1,\ldots,V_N\}, N = |\V|$ and a set of directed edges $\E \subseteq \V \times \V$. Let the feature vector associated with node $V_i$ be $\mathbf{x}_{i}$ which appears as a row in the node feature matrix $\X \in \mathbb{R}^{N\times m}$, where there are $m$ features per node. The blue agent thus observes $o \in \mathcal{O} = (\V,\E,\X)$ at each round.

As an example, consider the \cage environment, shown in \cref{fig:standard_defender}, with its state represented by \cref{table:challenge_2_state}. We transform this state into a graph representation where the vertices of the graph are the hosts and subnets and the edges in the graph represent the interfaces between the subnets. The resultant graph is shown in \cref{fig:cage_challenge_2_graph}.

\paragraph{Non-stationary \iit infrastructure}

An environment evolves as a consequence of agent actions, hence $P$ and $\Lambda$ model non-stationary distributions. $P$ and $\Lambda$ are fully determined by the connectivity of the network i.e. the topology of $G$. Due to the non-stationary nature of $G$, we adopt the following definition to describe a dynamic graph. The (static) graph representation of the \cage network is shown in \cref{fig:cage_challenge_2_graph}. Throughout we will use `graph' and `network' interchangeably to refer to the same object.

\begin{definition}[Non-stationary graph]
A non-stationary graph is a graph whose topology or properties change over time. Let $G_t = (\V_t, \E_t)$ represent the graph at time $t$, where: $\V_t$ is the set of vertices (or nodes) at time $t$ and $\E_t \subseteq \V_t \times \V_t$ is the set of edges (or links) at time $t$.
A non-stationary graph is then a time-dependent sequence of graphs $\G \triangleq \{ G_t \}_{t \in [T]}$, where $t\in [T]$ is the time domain and the following properties may change with time:
\begin{enumerate}[noitemsep,topsep=0pt]
    \item Vertices $\V_t$: nodes can appear, disappear or have attributes that change over time.
    \item Edges $\E_t$: edges can be created, removed or modified over time.
\end{enumerate}
The non-stationary nature of the topology refers to the dependence of the graph's structure and properties on time, such that $G_{t} \neq G_{t'}$ necessarily for some $t, t' \in [T]$.
\end{definition}
We seek a model which addresses two key challenges:
\begin{description}
	\item[Unpredictable topology changes] Real-world \textsc{it}/\ot networks experience constant topological shifts due to host activity, adversarial attacks and defensive actions. These changes create unpredictable dynamics that make training \acd agents difficult, as existing methods rely on fixed observation spaces or predefined defensive strategies.
	\item[Graph-based adaptation challenges] While graph representation learning techniques encode network states as graph embeddings for decision-making, they struggle with out-of-distribution generalization. Large, unforeseen topology shifts --such as subnet isolation due to attacks or defensive actions-- can significantly degrade agent performance.
\end{description}

\section{Related work}
\label{sec:related_work}

We focus on research that studies \emph{general} \acd (\gacd). There is a great deal of work that focuses on task-specific models, but few which can be carried between task in the general sense. To draw upon an analogy; we seek a model which, much like any software (e.g. antivirus), can be installed on any host or server and, with regular updates, requires minimal maintenance and can act autonomously to defend the target node. A model like that can handle dynamic network changes and a variety of attackers.

There exist many excellent review articles of recent \acd efforts, see e.g. the paper by \citet{vyas2023automated} but they are silent on the types of work we describe above. Recent work by \citet{nyberg2024structural} explores relational agent learning for automated incident response, where a message-passing neural network (\mpnn) agent learns relationships between network elements rather than specific network structures, allowing them to adapt to network changes without retraining. Their model cannot handle a diversity of red agents and is only trained against one type (`Meander' from \cage) contrasting with the \gacd approach which is designed to handle multiple types of red agents. To our knowledge, this is the only other work that explores \gacd.
\section{Method}
\label{sec:method}

We introduce the key components of our \emph{single-agent, multi-task} approach for \gacd{} -- an agent designed to defend against diverse attackers on previously unseen graph topologies. Three different variants of \gacd are presented: $\mathcal{M}_1, \mathcal{M}_2$ and $\mathcal{M}_3$.
\subsection{Embedding observations}
\label{sec:embedding}

Given a set of graph observations $\graphs$, we assume that topologies and states lie on a $d$-dimensional manifold, allowing projections into lower-dimensional spaces.  This aligns with the manifold distribution hypothesis (\mdh), which states that natural data distributions are concentrated on a low-dimensional manifold within a high-dimensional space \citep{lei2019geometric, tenenbaum2000global}. Learning this mapping enhances both understanding and reconstruction of the observation distribution. Once $\G$ has been successfully embedded, a blue agent can use this latent manifold as its state representation $s\in\mathcal{S}$, facilitating the development of a policy that accounts for transitions between different regions of the state space in response to changes in network topology. 

\begin{figure}[ht!]
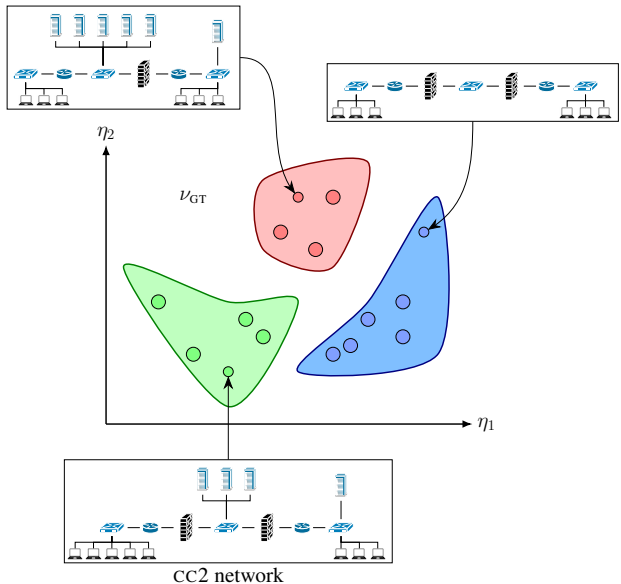

    \centering
    \resizebox{\columnwidth}{!}{
        \begin{tikzpicture}[-Latex, auto,every node/.style={font=\LARGE},]
    \draw[fill=green!25, draw=green!50!black,very thick] plot[smooth cycle] coordinates {(0, -1) (-3, 3) (0, 2) (2, 2) };
    \foreach \x/\y in {-2/2, -1/0.5, 1/1, 0.5/1.5}
    \draw[fill=green!50] (\x, \y) circle (2mm);
    \node [fill=green!50, draw=black, circle, inner sep=1mm] (cc2_node) at (0, 0) {};
    
    \draw[fill=red!25, draw=red!50!black, very thick] plot[smooth cycle] coordinates {(1, 5.5) (4, 6) (3, 3) (1, 3.5) };
    \foreach \x/\y in {1.5/4, 3/5, 2.5/3.5}
    \draw[fill=red!50] (\x, \y) circle (2mm);
    \node [fill=red!50, draw=black, circle, inner sep=1mm] (cc2_node_red) at (2, 5) {};
    
    \draw[fill=blue!50!cyan!50, draw=blue!50!black, very thick] plot[smooth cycle] coordinates {(2, 0) (4, 2) (6, 5) (6, 0.5) };
    \foreach \x/\y in {3/0.5, 3.5/0.75, 5/1, 4/1.5, 5/2}
    \draw[fill=blue!75!cyan!50] (\x, \y) circle (2mm);
    \node [fill=blue!75!cyan!50, draw=black, circle, inner sep=1mm] (cc2_node_blue) at (5.6, 4) {};

    \node (nu) at (-1,5) {$\nu_{\gt}$};
    
    \draw[-{Latex[scale=1]}, very thick] (-3.5,-1.5) -- (7,-1.5) node[right] {$\eta_1$};
    \draw[-{Latex[scale=1]}, very thick] (-3.5,-1.5) -- (-3.5,6.5) node[above] {$\eta_2$};

    \node (cc2) [scale=0.7, draw,label=below:\cage network] at (0,-4) {
      \input{content/figures/CC2_network_reduced.tex}
    };
    \node (reduced_cc2) [scale=0.7, draw] at (-3,9) {
      \input{content/figures/CC2_network_reduced_cascaded.tex}
    };
    \node (third_network) [scale=0.7, draw] at (7,8) {
      \input{content/figures/CC2_network_reduced_2}
    };

    \path [-{Stealth[scale=2]}] (cc2.north) edge (cc2_node);
    \path [-{Stealth[scale=2]}] (reduced_cc2.east) edge[out=0, in=130, looseness=1] (cc2_node_red);
    \path [-{Stealth[scale=2]}] (third_network.south) edge[out=-90, in=30, looseness=1] (cc2_node_blue);
\end{tikzpicture}
    }
    \caption{Projection of network topologies onto a low-dimensional latent manifold $\mathcal{Z} \in \mathbb{R}^d$ distributed as $\nu_{\gt}$. Each point represents a specific state of a network topology, projected onto the $\eta_1-\eta_2$ plane. The smaller circles represent the specifically shown embedded \iit infrastructures. Blobs of the same color correspond to states of the network of the same topology, clustered into distinct modes. The design of the latent space ensures that similar topologies are close, while topologies of different similarity appear further away on the manifold. Distribution $\nu_{\gt}$ describes the ground-truth distribution of this latent space.}
    \label{fig:2d_projection}
\end{figure}

We design the space such that all variants associated with this topology (i.e. the different states associated with it) lie in its neighborhood and can be clustered together, represented by modes of different color (or high-density `blobs') as shown in \cref{fig:2d_projection}.  Similarly, a different network topology can be projected onto a different region of the space such that all similar topologies are clustered around this region and so on and so forth. 

\paragraph{Example} To lend a less abstract understanding: consider a dataset of faces. While each face is unique, they might cluster into `modes' based on categories like gender, age group, ethnicity, lighting conditions or even just distinct facial expressions (e.g. smiling faces, neutral faces or angry faces). Our approach herein is analogous to this demarcation of the face space, but instead of faces we are interested in modes which describe the graph space $\mathcal{G}$.

A framework is required to embed $\graphs$.

\subsubsection{Graph encoding model}
\label{sec:encoding_model}

The variational graph autoencoder (\vgae) \citep{kipf2016variational} is a popular model for learning low-dimensional latent representation of graph properties such as nodes, edges and their features. Another graph embedding model is the Graphormer \citep{ying2021transformers} or graph transformer model (\gtm), which is an adaptation of the transformer architecture \citep{vaswani2017attention} specifically designed to process graph-structured information. 

The encoding model of the \gacd is designed as a 2-hop graph encoder operating with different graph convolution layers. Variants of our model alter the manner in which the node features are embedded prior to message passing, with $\mathcal{M}_{1}$ using a multilayer perceptron (\mlp) and $\mathcal{M}_{2}$ and $\mathcal{M}_{3}$ using transformers as the feature embedder. The initial feature vector for the nodes of the graph are designed based on knowledge of the \cage environment. The graph is comprised of three types of nodes, two static capturing the different subnets and hosts in the network topology and a dynamic node resulting from an action performed on a particular hhst, e.g. deploying a decoy. These node types are encoded as one-hot features of the nodes. In addition to these information about the state of a host is converted into a bit array and concatenated to the one-hot encoding of the node features. The state of a host in the network is encoded by its activity and compromised state, each of which can take four possible values. Hence, two 2-bit arrays are used to capture the state of the node, resulting in seven features for each node in the network. The edges of the graph are designed to follow the communication structure of the network as defined by the network control access lists, with undirected edges representing bi-directional communications and directed edges representing one-way interfaces. For decoy nodes deployed by the agent, we use a directed edge from the source host and another directed edge to the target host onto which the decoy is deployed. We model all communication interfaces as equivalent and hence do not include any edge features on the observation graph. Finally, the node embeddings are aggregated into graph embeddings using a mean pooling strategy.




The key innovation is that the model modifies the self-attention mechanism to incorporate graph structural information such as node degrees, edge distances and spatial encodings while maintaining the transformer's ability to capture long-range dependencies.


\subsubsection{Discontinuous latent space}

\begin{figure*}[ht!]
    \centering
    \begin{subfigure}[t]{0.49\textwidth}
        \centering
        \resizebox{0.85\columnwidth}{!}{
            \begin{tikzpicture}[-Latex,auto,every node/.style={font=\LARGE}]
    \draw[fill=green!25, draw=green!50!black, very thick] plot[smooth cycle] coordinates {(0, -1) (-3, 3) (0, 2) (2, 2) };
    \foreach \x/\y in {-2/2, -1/0.5, 1/1, 0.5/1.5}
    \draw[fill=green!50] (\x, \y) circle (2mm);
    \node [fill=green!50, draw=black, circle, inner sep=1mm] (cc2_node) at (0, 0) {};
    \draw[fill=red!50, opacity=0.75, draw=red!50!black, very thick] plot[smooth cycle] coordinates {(1, 5.5) (4, 6) (3, 3) (1, 3.5) };
    \foreach \x/\y in {1.5/4, 3/5, 2.5/3.5}
    \draw[fill=red!50] (\x, \y) circle (2mm);
    \node [fill=red!50, draw=black, circle, inner sep=1mm] (cc2_node_red) at (2, 5) {};
    \draw[fill=blue!50!cyan!50,opacity=0.75, draw=blue, very thick] plot[smooth cycle] coordinates {(2, 0) (4, 2) (6, 5) (6, 0.5) };
    \foreach \x/\y in {3/0.5, 3.5/0.75, 5/1, 4/1.5, 5/2}
    \draw[fill=blue!75!cyan!50] (\x, \y) circle (2mm);
    \node [fill=blue!75!cyan!50, draw=black, circle, inner sep=1mm] (cc2_node_blue) at (5.6, 4) {};
    \draw[-{Latex[scale=1]}, very thick] (-3.5,-1.5) -- (7,-1.5) node[right] {$\eta_1$};
    \draw[-{Latex[scale=1]}, very thick] (-3.5,-1.5) -- (-3.5,6.5) node[above] {$\eta_2$};

    \node [fill=red!50, draw=black, circle, inner sep = .5mm] (loc1) at (3,4.25) {$1$};
    \node [draw=black, circle, inner sep = .5mm] (loc2) at (4.5,4) {$2$};
    \node [draw=black, circle, inner sep = .5mm] (loc3) at (3.5,2.5) {$2$};
    \node [draw=black, circle, inner sep = .5mm] (loc4) at (2.4,1.5) {$3$};
    \node [draw=black, circle, inner sep = .5mm] (loc5) at (6,7) {$3$}
        node[left= 0.2cm of loc5] (node_label) {Outside observed distribution};

    \path [-{Stealth[scale=1.5]}, blue, thick] (cc2_node_red) edge (loc1);
    \path [-{Stealth[scale=1.5]}, red, thick] (loc1) edge (loc2);
    \path [-{Stealth[scale=1.5]}, green!50!black, thick] (loc1) edge (loc3);
    \path [-{Stealth[scale=1.5]}, blue, thick, dashed] (loc2) edge (loc5);
    \path [-{Stealth[scale=1.5]}, blue, thick, dashed] (loc3) edge (loc4);
\end{tikzpicture}
        }
        \caption{Illustration of an action that pushes the state (graph embedding) from known operating regions of the \grl agent. In the illustrated scenario, from an initial state, the blue agent takes an action that causes the state of the network to transition to a new position in the latent space -- see \circled{1}. From here, assume that either a red or green action \circled{2}, results in a transition to the state of the network such that the new state is outside the observed network topology data, \circled{3}.}
        \label{fig:2d_projection_with_actions}
    \end{subfigure}
    \hfill
    \begin{subfigure}[t]{0.49\textwidth}
        \centering
        \resizebox{0.7\columnwidth}{!}{
            \begin{tikzpicture}[-Latex,auto]

    \draw [very thick] (0, 0) coordinate(A) node[below left]{$0$} --
    (0, 6)coordinate(B) node[above left]{$1$} --
    (6, 6) coordinate(C) node[above right]{} --
    (6, 0) coordinate(D) node[below right]{$1$} --
    cycle;

    \node at (3, 0) [below] {$\overline{\eta_1}$};
    \node at (0, 3) [left] {$\overline{\eta_2}$};

    \draw [-,dashed,very thick] (B) -- (4, 3)coordinate(E);
    \draw [-,dashed,very thick] (A) -- (6, 4.5)coordinate(F);

    \coordinate (I) at (intersection of B--E and A--F);

    \fill[green!25,opacity=0.75](A)--(B)--(I)--cycle;
    \fill[red!50,opacity=0.75](B)--(I)--(F)--(C)--cycle;
    \fill[blue!50!cyan!50,opacity=0.75](A)--(D)--(F)--cycle;

    \foreach \x/\y in {0.3/5, 0.3/2, 1.5/4, 3/3}
    \draw[fill=green!50] (\x, \y) circle (2mm);
    \node [fill=green!50, draw=black, circle, inner sep=1mm] (cc2_node) at (0.25, 0.5) {};
    \foreach \x/\y in {1.5/5.5, 5.5/5.5, 4.75/4}
    \draw[fill=red!50] (\x, \y) circle (2mm);
    \node [fill=red!50, draw=black, circle, inner sep=1mm] (cc2_node_red) at (3.25, 5.5) {};
    \foreach \x/\y in {1.25/0.5, 3/0.5, 5.5/0.5, 3.5/2.25, 5.5/3.5}
    \draw[fill=blue!75!cyan!50] (\x, \y) circle (2mm);
    \node [fill=blue!75!cyan!50, draw=black, circle, inner sep=1mm] (cc2_node_blue) at (4.5, 1.5) {};

    \node [fill=red!50, draw=black, circle, inner sep = .5mm] (loc1) at (2.5,4.55) {$1$};
    \node [fill=blue!75!cyan!50, draw=black, circle, inner sep = .5mm] (loc2) at (2.8,1.25) {$2$};
    \node [fill=blue!75!cyan!50, draw=black, circle, inner sep = .5mm] (loc3) at (5,2.5) {$2$};
    \node [fill=green!50, draw=black, circle, inner sep = .5mm] (loc4) at (1,2) {$3$};
    \node [fill=red!50, draw=black, circle, inner sep = .5mm] (loc5) at (4,4.5) {$3$};

    \path [-{Stealth[scale=1.5]}, blue, thick] (cc2_node_red) edge (loc1);
    \path [-{Stealth[scale=1.5]}, red, thick] (loc1) edge (loc2);
    \path [-{Stealth[scale=1.5]}, green!50!black, thick] (loc1) edge (loc3);
    \path [-{Stealth[scale=1.5]}, blue, thick, dashed] (loc2) edge (loc4);
    \path [-{Stealth[scale=1.5]}, blue, thick, dashed] (loc3) edge (loc5);

\end{tikzpicture}
        }
        \caption{Illustration of actions which push the state around known operation regions of the \grl agent, in the $\overline{\eta_1} - \overline{\eta_2}$ plane. An initial blue action causes a transition to \circled{1}. From here a red action moves the state to the blue (known) operating region, a green action has the same effect -- see \circled{2}. The blue agent can then react which results in a state transition, again, to known operation regions, guaranteeing a valid defensive action.}
        \label{fig:2d_projection_with_ot}
    \end{subfigure}
    \caption{Illustrations of state (graph embedding) transitions in the latent space resulting from agent actions. \Cref{fig:2d_projection_with_actions} illustrates a scenario wherein following an action by a blue agent, subsequent actions by the red or green agents can transition the state to positions outside the observed data distribution. \cref{fig:2d_projection_with_ot} depicts actions that keep the network state within known operating regions, illustrating how the blue agent’s defensive actions can ensure valid state transitions within the latent space.}
\end{figure*}
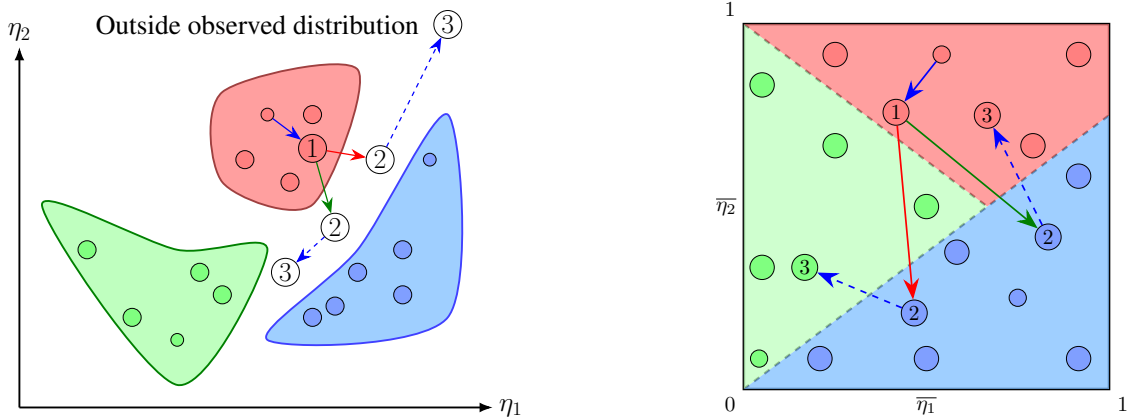

A key challenge with our approach lies in ensuring that the learned projections of network states align with the distribution of the observed data, as discrepancies (i.e. discontinuities) could prevent effective decision-making. Consider the scenario illustrated in \cref{fig:2d_projection_with_actions} where, from an initial state, the blue agent takes an action that causes the state of the network to transition to a new position in the latent space -- see \circled{1} in \cref{fig:2d_projection_with_actions}. From here, in the next round, assume that either a red or green action, \circled{2} in \cref{fig:2d_projection_with_actions}, results in a transition to the state of the network such that the new state is outside the observed network topology data, \circled{3}. In such a scenario, a graph representation learning agent (or \emph{blue} agent) cannot provide guarantees for the defense of a network as the new observation embedding lies in a region of space that it has not been trained to defend. Instead, any further action by the blue agent can result in further deviations of the state of the topology away from observed regions.

\subsection{Objective} 
\label{sec:objective}

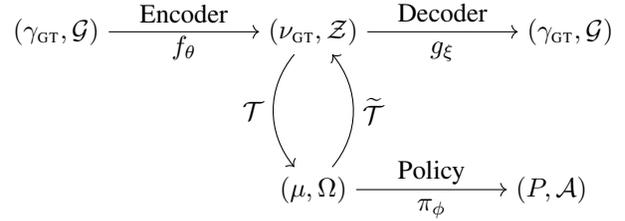
\begin{figure}[H]
    \centering
    \begin{tikzpicture}[node distance = 2cm]
        \node (data_in) [] at (-2,0) {$(\gamma_{\gt}, \mathcal{G})$};
        \node (latent_space) [right = of data_in] {$(\nu_{\gt}, \mathcal{Z})$};
        \node (noise_distro) [below = 1.5cm of latent_space] {$(\mu, \Omega)$};
        \node (data_out) [right = of latent_space] {$(\gamma_{\gt}, \mathcal{G})$};
        \node (action_space) [right = of noise_distro] {$(P,\mathcal{A})$};

        \draw [->] (data_in) to node[align=center] {Encoder \\ $f_{\theta}$} (latent_space);
        \draw [->] (latent_space) to node[align=center] {Decoder \\ $g_{\xi}$} (data_out);
        \draw [->] (latent_space) to [bend right= 40] node[midway,left] {$\mathcal{T}$} (noise_distro);
        \draw [->] (noise_distro) to [bend right= 40] node[midway,right] {$\widetilde{\mathcal{T}}$} (latent_space);

        \draw [->] (noise_distro) to node[align=center] {Policy \\ $\pi_{\phi}$} (action_space);
    \end{tikzpicture}
    \caption{Commutative diagram showing the constituent objects and morphisms required to train a blue agent policy within the encoder-decoder-\ot framework. In each tuple the distribution is the first argument followed by the domain.
    }
    \label{fig:commutative_diagram}
\end{figure}

To deal with the challenges discussed in the previous section, we construct objectives that rely on the \mdh being true (see the start of \cref{sec:embedding}). In encoder-decoder-based \gnn{}s, the generator’s primary objective is to approximate the true data distribution, ensuring that its generated samples are located in the lower-dimensional data manifold. Let $\G \subset \mathcal{G}$ be distributed as $\gamma_{\gt}$, the true data distribution, where $\mathcal{G}$ is the set of all undirected graphs. The encoder $f_{\theta}$ is trained to encode the data manifold from $\mathcal{G}$ to the latent space $\mathcal{Z}$ distributed as $\nu_{\gt}$.  The decoder $g_{\xi}$ samples from the latent distribution $\nu_{\gt}$ and maps these back to the data manifold $\mathcal{G}$, see \cref{fig:commutative_diagram}. 
The distributions used in \cref{fig:commutative_diagram}, as well in the rest of this section, are explained below.
\begin{description}[font=\normalfont]
    \item[$\nu_{\gt}$] Underlying ground-truth distribution of data in the latent space $\mathcal{Z}$ (theoretical and unknown)
    \item[$\widehat{\nu}_{\gt}$] The empirical distribution of latent codes $\Z$ obtained by encoding the finite set of graphs (the actual discrete target for \sdot, defined overleaf) 
    \item[$\nu$] Learned continuous latent distribution, constructed by pushing forward a simple noise distribution through the extended \sdot map 
    \item[$\mu$] Uniform noise distribution $\mathcal{U}([0,1]^d)$ supported on $\Omega \triangleq [0,1]^d$
\end{description}

To get a handle on the meaning and purpose of these distributions, we have provided an additional analogy in \cref{sec:map_analogy}.

\paragraph{Mode collapse and mixture} The encoder-decoder approach described in \cref{fig:commutative_diagram} is well understood but suffers from \emph{mode collapse} and \emph{mode mixture} \citep{an2019ae, nagarajan2017gradient}. Mode collapse occurs when the generator fails to capture the diversity of $\gamma_{\gt}$, producing limited or repetitive outputs by focusing only on a few high-density regions while ignoring others. This leads to a disconnected or sparse latent space (see e.g. \cref{fig:2d_projection_with_actions} where the domain is not fully covered by the colored sub-spaces), where large portions of the input map to the same or very few outputs, creating discontinuities that result in abrupt jumps between generated samples rather than smooth transitions. In contrast, mode mixture leads to unrealistic interpolations between different data modes, causing blurry boundaries in the latent space. Instead of forming well-separated clusters, the latent representations overlap, blending multiple modes and failing to preserve distinct categories. Essentially, generative models struggle with mode collapse and mixture because they try to use neural networks to estimate mappings that are fundamentally discontinuous. 


\subsection{Optimal transport to the rescue} 
\label{sec:optimal_transport}

Optimal transport (\ot) \citep{solomon2018optimal} has been suggested as a way of dealing with mode collapse and mode mixture. For example, \citet{an2019ae} introduce the \aeot model (top part of \cref{fig:commutative_diagram}) to deal with the above issues. They use an autoencoder (\textsc{ae}) to embed images (we embed graphs) in a low-dimensional manifold, while preserving important features upon which they employ optimal transport to map from a random noise distribution $\mu$, to the latent manifold $\mathcal{Z}$, in a way that matches the distribution of the embedded real images. 
We use similar ideas to embed attributed graphs $\G$, and treat the resulting latent manifold as the state-space $\mathcal{S}$ in a \pomdp (see policy morphism in \cref{fig:commutative_diagram}).

As seen in \cref{fig:commutative_diagram}, we employ two transport maps $\tmap$ and $\widetilde{\tmap}$ to both map into the noise distribution $\mu$ as well as out of it. This allows us to use the samples in the noise distribution as a representation for our observations by mapping an observed sample from latent space to the noise space with $\tmap$. As this space is structured to accurately model the observed samples, we can ensure that the resultant embedding would be an experience that the agent is trained to defend.

\paragraph{Regular transport map} Within the latent manifold $\mathcal{Z}$ in \cref{fig:commutative_diagram}, latent codes $\Z = \{\z_i \in \mathbb{R}^d \mid i \in [T]\}$ distributed as $\widehat{\nu}_{\gt}$, will typically cluster into different modes. Latent codes are mapped to $\mu$ using transport map 
\begin{equation}
    \tmap:\widehat{\nu}_{\gt} \xrightarrow{h_{\psi}} \mu
\end{equation}
shown on the left in \cref{fig:commutative_diagram}. We let the transformation function $h_{\psi}$ be a feedforward neural network \citep{mcculloch1943logical} with parameters $\psi$.  


\paragraph{Semi-discrete optimal transport map} Following \citet{an2019ae, an2020ae}, we seek a continuous latent distribution $\nu$ as seen in \cref{fig:2d_projection_with_ot} -- i.e. a latent space without discontinuities. To be effective, $\nu$ must fulfill a set of desiderata \citep{an2019ae}:
\begin{enumerate}
    \item $\nu$ has to properly generalize the true empirical latent distribution $\widehat{\nu}_{\gt}$ so that all modes are covered by its support \citep{an2020ae}.
    \item The support of $\nu$ must have a topology similar to that of the ground truth $\nu_{\gt}$ to ensure that the map from $\nu$ to $\gamma$ is continuous.
    \item $\nu$ has to be efficient to sample from.
\end{enumerate}
The semi-discrete optimal transport (\sdot) map $\widetilde{\tmap}$ fulfills all desiderata. Taking our cue from the original work by \citet{an2019ae} we expose all steps that eventually lead to $\widetilde{\tmap}$. We begin by making the observation that the target of the transportation is a discrete set of points $\Z$. The corresponding target measure, $\nu$, is represented as a Dirac measure for each discrete point: $\nu = \sum_{i=1}^T \nu_i \delta(\z - \z_i)$. Here, $\nu_i$ represents the mass or probability associated with each point $\z_i$. Crucially, the \emph{total mass} of the source and target measures must be equal: $\mu(\Omega) = \sum_{i=1}^n \nu_i$.

When a semi-discrete transport map $\tau: \Omega \to \Z$ is applied, it naturally induces a decomposition of the source domain $\Omega$ into distinct cells $\{W_1, W_2, \dots, W_T\}$, such that $\Omega = \bigcup_{i=1}^T W_i$. Within this setup, every point $x$ located in a specific cell $W_i$ is mapped directly to its corresponding target point $\z_i$. This means the map $\tau$ effectively assigns all points within $W_i$ to $\z_i$.

For the map $\tau$ to be considered measure-preserving (denoted $\tau_\# \mu = \nu$), the $\mu$-volume (or probability mass) of each cell $W_i$ must be equal to the $\nu$-measure (mass) of its image, $\tau(W_i) = \z_i$. The cost function \citep[\S 7.2.2]{santambrogio2015optimal} is given by $c : \Omega \times \Z \to \mathbb{R}$ where $c(x,\z)$ represents the cost of transporting a unit mass from $x$ to $\z$. The total cost of $\tau$ is given by 
\begin{equation}
    \label{eq:total_ot_loss}
    \int_{\Omega} c(x,\tau(x))\textrm{d}\mu(x) = \sum_{i=1}^n \int_{W_i} c(x,\z_i)\textrm{d}\mu(x).
\end{equation}
A common choice for $c$ is the Euclidean distance, but our setting demands that we use a distance more suited to measuring the distance between two graphs $G_i$ and $G_j$. We employ the Fused Gromov-Wasserstein (\fgw) distance \citep{vayer2018optimal} which measures similarities between two attributed graphs (see \cref{sec:fgw_distance} for more details). The \sdot is the measure-preserving map that minimizes the total cost: 
\begin{equation*}
    \tau^* =\argmin_{\tau_{\#\mu} = \nu} \int_{\Omega} c(x, \tau(x))\textrm{d}\mu(x) 
\end{equation*}
The map $\tau^*$ is \emph{piece-wise linearly extended to a global continuous map} $\widetilde{\tmap}$ \citep{an2019ae}. The details can be found in the original paper (incl. optimization details) but we will dwell on the key idea here owing to its importance to this work. While $\tau^*$ maps an entire cell $W_i$ to a single point $\z_i$, $\widetilde{\tmap}$ takes the latent codes and the structure of the cells $W_i$ to create a continuous mapping -- by filling in the `gaps'.

The gaps are filled by considering the latent codes, they are connected to form a mesh or simplicial complex\footnote{A simplicial complex is the result of tiling or approximating a space using triangles and tetrahedrons (in two and three dimensions respectively), in a way that preserves topological properties like holes, connectivity and boundaries.}. As the simplicial space preserves topological properties, the singularity set in the source domain $\mathcal{G}$ can be located and avoided when generating new graphs. Consequently, random noise from $\Omega$ can be used in conjunction with $g_{\xi}$, to generate new and valid graphs through $g_{\xi} \circ \widetilde{\tmap}$. Simpler still: $\tau^*$ finds the optimal way to distribute continuous noise to the specific discrete latent modes. It tells us which regions of the noise map to which latent mode. $\widetilde{\tmap}$ builds on $\tau^*$; it takes the optimal cell decomposition and the latent codes and extends these into a continuous piecewise linear map. This allows for generating a continuum of diverse samples from the noise distribution, avoiding abrupt jumps at mode boundaries and provides a direct map from any random noise input to a meaningful latent code for generation.


\subsection{Synthesizing a non-stationary environment}
\label{sec:body_psg}

The set $\graphs$ represents the graph topologies of a non-stationary environment -- $G_t$ may change as $t$ evolves. A priori, we do not have direct access to these topologies. However, we assume knowledge of the magnitude  $N$  of the target infrastructure (i.e., the number of nodes), which allows us to naively generate a large number of undirected graphs\footnote{There are a total of $2^{\frac{N(N-1)}{2}}$ undirected graphs in $N$ vertices.} using e.g. the Erd\H{o}s--R\'enyi model \citep{erdos1960evolution}.

Often, we can leverage domain-specific knowledge to better understand how the topology may evolve based on its design and the functionality of the attached nodes. This knowledge allows us to synthesize more accurate undirected graphs that are generated according to environmental rules. To achieve this, we introduce a procedural scenario generator (\psg), described in \cref{alg:procedural_scenario_generator} in the supplement, which generates scenarios resembling those in the \cage environment illustrated in \cref{fig:standard_defender}. Examples of generated networks are shown in \cref{fig:2d_projection}. For further details, refer to \cref{sec:psg} of the supplement.



\subsection{Training the blue agent}
\label{sec:agent_training}

As noted at the start of this section, we present three model variations, indexed by the terms in the loss function.  
\begin{description}[font=\normalfont]
    \item[$\mathcal{M}_1$: $\loss_{\mse} + \loss_{\fgw} + \loss_{\ppo}$]
    A modified \mpnn implementation that enforces \ot on the latent space. The encoding model is trained end-to-end with the \ppo agent \citep{schulman2017proximal}, with both the reward and transformation errors arising from $\tmap$ and $\widetilde{\tmap}$ incorporated into the training loss.
    
    \item[$\mathcal{M}_2$: $\loss_{\mse} + \loss_{\fgw} + \loss_{\ppo}$]
    Implements a Graphormer \citep{ying2021transformers} or graph transformer model (\gtm) architecture to improve the expressiveness of the agent, augmenting it with the \ot mapping. In this case, the \gtm model is trained end-to-end with the \ppo agent, with both the reward and transformation errors from $\tmap$ and $\widetilde{\tmap}$ incorporated into the training loss.
    
    \item[$\mathcal{M}_3$: $\loss_{\textsc{ae}} + \loss_{\fgw} + \loss_{\mse}; \loss_{\ppo}$]
    Uses the \gtm model in an auto-encoding setup which is first pre-trained using the sum of the \ot cost associated with the transformation $\widetilde{\tmap}$, the \mse loss associated with the transformation $\tmap$, and the reconstruction loss $\loss_{\textsc{ae}}$ associated with the graph samples, as formulated by \citet[Equation 3]{kipf2016variational}. Once trained, the encoding layers of the \gtm are used with the action network to construct the \gacd agent. In this case, the \gtm encoder's weights are frozen, with the \ppo agent being trained using the clipped loss formulation of $\loss_{\ppo}$ as formulated by \citet[Equation 7]{schulman2017proximal}.
\end{description}

With that we can write down losses associated with the \ot segment of \cref{fig:commutative_diagram}
\begin{align}
    \loss_{\mse} &= \sum_{i=1}^n = \| h_{\psi}(\z_i) - x_i\|_2 \\
    \loss_{\fgw} &= \frac{1}{n} \sum_{i=1}^n c(x_i,\tau(x_i)) \label{eq:fgw_loss}
\end{align}
where \cref{eq:fgw_loss} corresponds to the Monte-Carlo approximation of \cref{eq:total_ot_loss}.

\begin{figure*}[ht!]
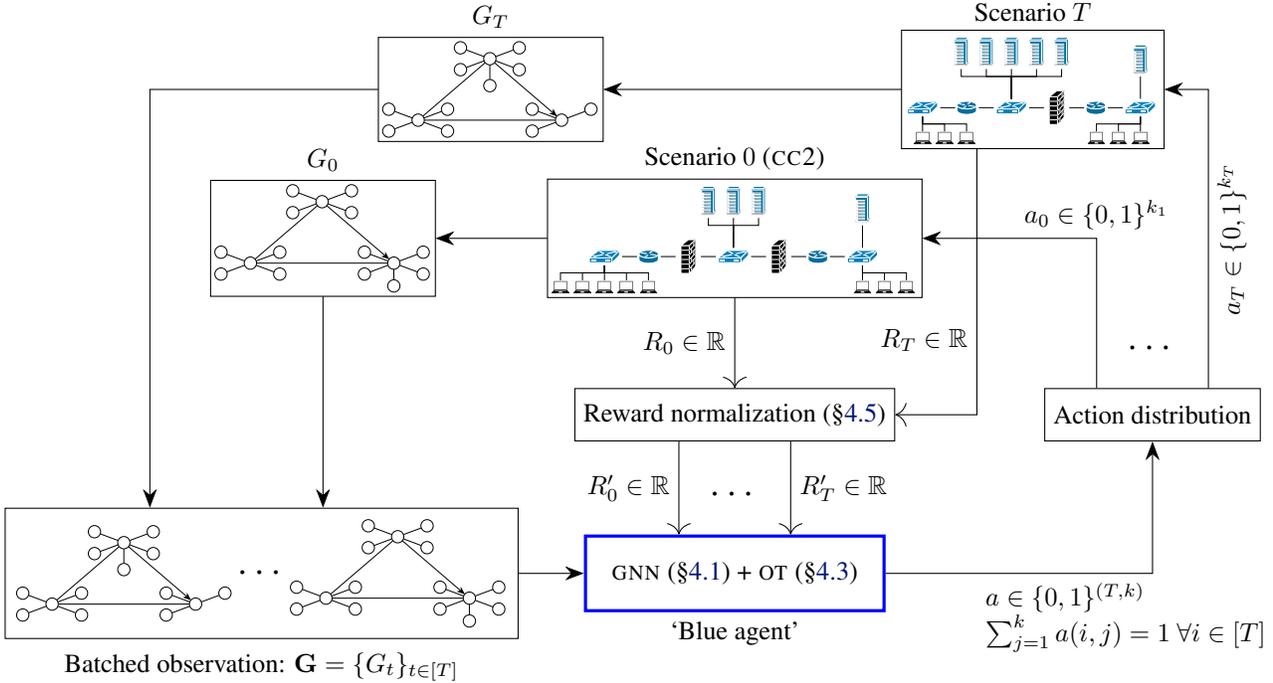

    \centering
    \resizebox{\textwidth}{!}{
        \begin{tikzpicture}[node distance = 1cm]

    \tikzset{process/.style={
        rectangle,
        draw=black,
        fill=white,
        align=center,
        minimum width=4em,
        minimum height=2em
    },
    arrow/.style={-{Stealth[scale=1.5]}}
    }

    \node (blue_agent) [draw=blue, very thick, inner sep=0.35cm,label=below:`Blue agent'] at (0, -0.5) {\gnn (\cref{sec:embedding}) +  \ot (\cref{sec:optimal_transport})};
    \node (norm) [process, above= 1.25cm of blue_agent]  {Reward normalization (\cref{sec:agent_training})};
    \node (action) [process, right=  2cm of norm]  {Action distribution};
    \node (hdots) [above= 0.3cm of action]  {\Large$\cdots$};
    \node [above= 0.3cm of blue_agent] (more_dots) {\Large$\cdots$};

    \node (cc1) [scale=0.4, draw,label=above:Scenario $0$ (\textsc{cc2})] at (0,4) {
      \input{content/figures/CC2_network_reduced.tex}
    };
    \node (cc2) [scale=0.4, draw,label=above:Scenario $T$] at (4,6) {
      \input{content/figures/CC2_network_reduced_cascaded.tex}
    };

    \node (cc1_graph) [scale=0.275, draw, inner sep=0.2cm, label=above:$G_0$, left = 1.5cm of cc1] {
      \begin{tikzpicture}[
    node distance=10mm,
    my_node/.style={circle, draw, minimum size=6mm},
    arrow/.style={-{Stealth[scale=1.5]}}
]

\node [my_node, label={[label distance=0mm]above:{}}] at (0,0) (os) {};
\node [my_node,above left = 1mm and 10mm of os, ] (os1) {};
\node [my_node,below left = 1mm and 10mm of os,] (os2) {};
\node [my_node,above right= 1mm and 10mm of os,] (os3) {};
\node [my_node,below right= 1mm and 10mm of os,] (os4) {};
\foreach \i in {os1, os2, os3, os4} {
    \draw (os) --  (\i);
}

\node [my_node, label={[label distance=0mm]above:{}}] at (-3.5,-3) (es) {};
\node [my_node,above left = 1mm and 10mm of es,] (es1) {};
\node [my_node,below left = 1mm and 10mm of es,] (es2) {};
\node [my_node,above right= 1mm and 10mm of es,] (es3) {};
\node [my_node,below right= 1mm and 10mm of es,] (es4) {};
\foreach \i in {es1, es2, es3, es4} {
    \draw (es) --  (\i);
}

\node [my_node, label={[label distance=0mm]above:{}}] at (3.5,-3) (us) {};
\node [my_node,above left = 1mm and 10mm of us,] (us1) {};
\node [my_node,below left = 1mm and 10mm of us,] (us2) {};
\node [my_node,above right= 1mm and 10mm of us,] (us3) {};
\node [my_node,below right= 1mm and 10mm of us,] (us4) {};
\node [my_node,below = 5mm of us,] (us5) {};
\foreach \i in {us1, us2, us3, us4, us5} {
    \draw (us) --  (\i);
}

\draw[arrow, thick] (os) -- (us);
\draw[thick] (os) -- (es);
\draw[thick] (es) -- (us);

\end{tikzpicture}
    };
    \node (cc2_graph) [scale=0.275, draw,inner sep=0.2cm, label=above:$G_T$, left= 4cm of cc2] {
      \begin{tikzpicture}[
    node distance=10mm,
    my_node/.style={circle, draw, minimum size=6mm},
    arrow/.style={-{Stealth[scale=1.5]}}
]

\node [my_node, label={[label distance=0mm]above:{}}] at (0,0) (os) {};
\node [my_node,above left = 1mm and 10mm of os, ] (os1) {};
\node [my_node,below left = 1mm and 10mm of os,] (os2) {};
\node [my_node,above right= 1mm and 10mm of os,] (os3) {};
\node [my_node,below right= 1mm and 10mm of os,] (os4) {};
\node [my_node,below = 7mm of os] (os5) {};
\foreach \i in {os1, os2, os3, os4, os5} {
    \draw (os) --  (\i);
}

\node [my_node, label={[label distance=0mm]above:{}}] at (-3.5,-3) (es) {};
\node [my_node,above left = 1mm and 10mm of es,] (es1) {};
\node [my_node,below left = 1mm and 10mm of es,] (es2) {};
\node [my_node,above right= 1mm and 10mm of es,] (es3) {};
\node [my_node,below right= 1mm and 10mm of es,] (es4) {};
\foreach \i in {es1, es2, es3, es4} {
    \draw (es) --  (\i);
}

\node [my_node, label={[label distance=0mm]above:{}}] at (3.5,-3) (us) {};
\node [my_node,above left = 1mm and 10mm of us,] (us1) {};
\node [my_node,below left = 1mm and 10mm of us,] (us2) {};
\node [my_node,above right= 1mm and 10mm of us,] (us3) {};
\foreach \i in {us1, us2, us3} {
    \draw (us) --  (\i);
}

\draw[arrow, thick] (os) -- (us);
\draw[thick] (os) -- (es);
\draw[thick] (es) -- (us);

\end{tikzpicture}
    };

    \node (hg1) [scale=0.275, left = 1cm of blue_agent] {
      \begin{tikzpicture}[
    node distance=10mm,
    my_node/.style={circle, draw, minimum size=6mm},
    arrow/.style={-{Stealth[scale=1.5]}}
]

\node [my_node, label={[label distance=0mm]above:{}}] at (0,0) (os) {};
\node [my_node,above left = 1mm and 10mm of os, ] (os1) {};
\node [my_node,below left = 1mm and 10mm of os,] (os2) {};
\node [my_node,above right= 1mm and 10mm of os,] (os3) {};
\node [my_node,below right= 1mm and 10mm of os,] (os4) {};
\foreach \i in {os1, os2, os3, os4} {
    \draw (os) --  (\i);
}

\node [my_node, label={[label distance=0mm]above:{}}] at (-3.5,-3) (es) {};
\node [my_node,above left = 1mm and 10mm of es,] (es1) {};
\node [my_node,below left = 1mm and 10mm of es,] (es2) {};
\node [my_node,above right= 1mm and 10mm of es,] (es3) {};
\node [my_node,below right= 1mm and 10mm of es,] (es4) {};
\foreach \i in {es1, es2, es3, es4} {
    \draw (es) --  (\i);
}

\node [my_node, label={[label distance=0mm]above:{}}] at (3.5,-3) (us) {};
\node [my_node,above left = 1mm and 10mm of us,] (us1) {};
\node [my_node,below left = 1mm and 10mm of us,] (us2) {};
\node [my_node,above right= 1mm and 10mm of us,] (us3) {};
\node [my_node,below right= 1mm and 10mm of us,] (us4) {};
\node [my_node,below = 5mm of us,] (us5) {};
\foreach \i in {us1, us2, us3, us4, us5} {
    \draw (us) --  (\i);
}

\draw[arrow, thick] (os) -- (us);
\draw[thick] (os) -- (es);
\draw[thick] (es) -- (us);

\end{tikzpicture}
    };
    \node (hg2) [scale=0.275, left = 0.7cm of hg1,label={[label distance=-0.1cm]right:{\Large $\cdots$}}] {
      \begin{tikzpicture}[
    node distance=10mm,
    my_node/.style={circle, draw, minimum size=6mm},
    arrow/.style={-{Stealth[scale=1.5]}}
]

\node [my_node, label={[label distance=0mm]above:{}}] at (0,0) (os) {};
\node [my_node,above left = 1mm and 10mm of os, ] (os1) {};
\node [my_node,below left = 1mm and 10mm of os,] (os2) {};
\node [my_node,above right= 1mm and 10mm of os,] (os3) {};
\node [my_node,below right= 1mm and 10mm of os,] (os4) {};
\node [my_node,below = 7mm of os] (os5) {};
\foreach \i in {os1, os2, os3, os4, os5} {
    \draw (os) --  (\i);
}

\node [my_node, label={[label distance=0mm]above:{}}] at (-3.5,-3) (es) {};
\node [my_node,above left = 1mm and 10mm of es,] (es1) {};
\node [my_node,below left = 1mm and 10mm of es,] (es2) {};
\node [my_node,above right= 1mm and 10mm of es,] (es3) {};
\node [my_node,below right= 1mm and 10mm of es,] (es4) {};
\foreach \i in {es1, es2, es3, es4} {
    \draw (es) --  (\i);
}

\node [my_node, label={[label distance=0mm]above:{}}] at (3.5,-3) (us) {};
\node [my_node,above left = 1mm and 10mm of us,] (us1) {};
\node [my_node,below left = 1mm and 10mm of us,] (us2) {};
\node [my_node,above right= 1mm and 10mm of us,] (us3) {};
\foreach \i in {us1, us2, us3} {
    \draw (us) --  (\i);
}

\draw[arrow, thick] (os) -- (us);
\draw[thick] (os) -- (es);
\draw[thick] (es) -- (us);

\end{tikzpicture}
    };
    \node [fit=(hg1)(hg2), draw] (hypergraph) {};
    \node [below = 0.05cm of hypergraph] (hypergraph_label) {Batched observation: $\G = \graphs$};

    \draw [arrow] (blue_agent.east) -| node[pos=0.45, below, align=left] {$a \in \{0, 1\}^{(T, k)}$ \\ $\sum_{j=1}^{k}a(i, j)=1 \ \forall i \in [T]$} (action.south);
    
    \draw [arrow] ([xshift=-0.75cm]action.north) |- node[pos=0.5, above] {$a_0 \in \{0, 1\}^{k_1}$} (cc1.east);
    \draw [arrow] ([xshift=0.75cm]action.north) |- node[rotate=90, below, pos=0.25] {$a_T \in \{0, 1\}^{k_T}$} (cc2.east);
    \draw [-{>[scale=1.75]}] (cc1.south) --  node[midway, left] {$R_0 \in \mathbb{R}$} ([xshift=0] cc1.south |- norm.north);
    
    \draw [-{>[scale=1.75]}] ([xshift=-0.75cm]cc2.south) |- node[pos=0.36,left] {$R_T \in \mathbb{R}$} (norm.east);
    
    \draw [-{>[scale=1.75]}] ([xshift=-0.75cm]norm.south) --  node[midway, left] {$R_0' \in \mathbb{R}$} ([xshift=-0.75cm] norm.south |- blue_agent.north);
    \draw [-{>[scale=1.75]}] ([xshift=0.75cm]norm.south) --  node[midway, right] {$R_T' \in \mathbb{R}$} ([xshift=0.75cm] norm.south |- blue_agent.north);

    \draw [arrow] (cc1.west) -- (cc1_graph.east);
    \draw [arrow] (cc2.west) -- (cc2_graph.east);

    \draw [arrow] (cc1_graph.south) -- ([xshift=0] cc1_graph.south |- hypergraph.north);
    \draw [arrow] (cc2_graph.west) -| ([xshift=-1.5cm]hypergraph.north);

    \draw [arrow] (hypergraph.east) -- (blue_agent.west);

\end{tikzpicture}
    }
    \caption{Illustration of the single-agent multi-task setting in which the \gacd agent is trained. We randomly sample a set of environments, Scenario $0$ through Scenario $T$, by varying the topology using the \psg{} -- see \cref{alg:procedural_scenario_generator}. The environment's observations are represented as graphs that are aggregated into a batched observation $\G$ that is fed to the blue agent. The agent consists of a \gnn encoder and an \ot map that projects the graph onto the domain $\Omega$ from which an action network predicts a distribution of actions for each environment, with each having $k$ possible actions. The predicted actions are distributed to each environment as appropriate. Rewards, $R_i$, for the different environments are normalized to the baseline \cage range prior to being fed to the agent.}
    \label{fig:blue_agent}
\end{figure*}

The blue agent is trained as shown in \cref{fig:blue_agent}, in a multi-task setting with each sampled environment representing a new task for the agent to learn. These environments are generated by the \psg with a random red agent, two types of red agents are available in the \cage scenario, assigned to each task. The blue agent operates on a batch of observations represented as a batched graph, each of which is projected into the random noise space $\mu$, with the transformation map $\tmap$. The embeddings in this uniform noise space represent the observations passed to the \ppo algorithm which is comprised of two action networks, one identifying the host node which is acted upon by the blue agent and the second which identifies the action to be performed by the blue agent on the selected host node. The action that can be performed by the agent is given by 
\begin{equation}
    \begin{aligned}
    n_i &\sim P[h_1(x_i \mid \alpha)] \\
    a_n &\sim P[h_2(x_i \mid \beta, n_i)]
    \end{aligned}
\end{equation}
where $h_1(\cdot \mid \alpha):\mathbb{R}^{d} \to \mathbb{R}^1$ represents an action network predicting energies $e_i$ over the nodes of the network; $h_2(\cdot \mid \beta, n_i):\mathbb{R}^{d} \to \mathbb{R}^k$ represents a another action network that learns a distribution over the actions of the selected host. With the estimated energies, the probability of selecting a node that is acted upon is computed using a masked softmax operation. 

The training is carried out for $2 \times 10^6$ time-steps, across ca. $20,000$ episodes. We train the agent on episodes having a maximum of $10$0 step length. To accommodate errors in the agent’s actions that result in failures to the simulation, we penalize such actions with a large negative reward $(-1500)$ truncating the episode when such failure is encountered.  While training time is higher than the baseline Cardiff \ppo model \citep{vyas2023automated}, the training time per epoch is inline with other \gnn models. Note, in contrast to existing methods, our agent only needs to be trained \emph{once} as it learns to generalize across network topologies. The inference time of all our agents are  in the same order of magnitude as other \ppo agents.

Note that $\mathcal{M}_1$, $\mathcal{M}_2$ and $\mathcal{M}_3$ do not interact with each other and are trained independently, with a fixed set of topologies. $\mathcal{M}_1$ and $\mathcal{M}_2$ share the same training procedure, while $\mathcal{M}_2$ and $\mathcal{M}_3$ share the same parameter count. As the models do not interact with each other, there are no confounding factors across them.

\paragraph{Reward normalization procedure} The reward for the different environments are normalized to the baseline \cage range, prior to being fed to the blue agent. We explored different normalization schemes but having analyzed the results, we concluded that it is not possible to compare the two settings. For example, a topology that has three operational servers can have three times worse reward than an topology with a single operational server, as each impact on the operational server can lead to a high penalty. Hence, there is a need to normalize the rewards to the same scale -- chosen as described to the \cage range.


To train the agents, we combine the representation learning step, with that of training the agent in a \ppo setting, optimizing for the losses as described in previous section.

\section{Experiments}
\label{sec:experiments}

We evaluate\footnote{A Python implementation is available. Please contact AR for access.} the \gacd models in a series of experiments:
\begin{itemize}
    \item[\cref{sec:scalability}] Evaluate the scalability and generalizability of the model. 
    \item[\cref{sec:ablation_study}] Ablation study.
    \item[\cref{sec:scaling_to_larger_networks}] We increase the size of the \cage network and evaluate performance.
    \item[\cref{sec:novel_pattern}] We study novel attack patterns.
    \item[\cref{sec:comparison_sota}] Compare \gacd's performance against the \mpnn model from \citet{nyberg2024structural}.
    \item[\cref{sec:randomization}] Compare performance against the \sota \ppo variant of the model under varying observation conditions.
    \item[\cref{sec:dynamic_env}] Evaluate the performance of the model in a setting where the network topology is altered midway through the episode.
    \item[\cref{sec:optimal_transport_utility}]  Finally we remove the \ot layer to measure its overall utility.
\end{itemize}

\subsection{Model scalability}
\label{sec:scalability}

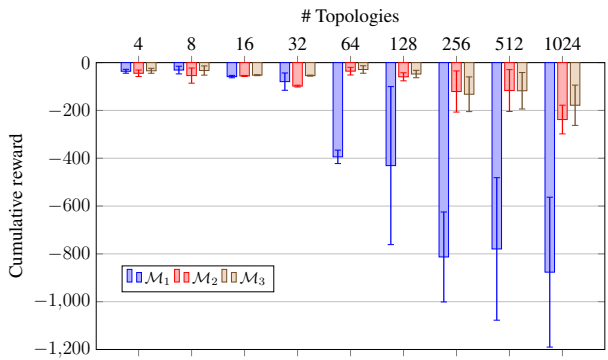
\begin{figure}[ht!]
    \centering
    \resizebox{\columnwidth}{!}{
        \pgfplotsset{compat=newest}
\pgfplotsset{/pgfplots/error bars/error bar style={thick}}
\begin{tikzpicture}
\begin{axis}[
    width=0.75\textwidth,
    height=0.75\textwidth / 1.618,
    tick label style={font=\large},
    label style={font=\large},
    ymajorgrids,
    ybar,
    bar width=6pt,
    enlarge x limits=0.1,
    ymin=-1200, ymax=0,
    xlabel={\# Topologies},
    ylabel={Cumulative reward},
    symbolic x coords={4, 8, 16, 32, 64, 128, 256, 512, 1024},
    xticklabel pos=upper, 
    xtick=data,
    legend style={at={(0.2,0.2)},anchor=south,legend columns=-1}, 
    legend entries={$\mathcal{M}_1$, $\mathcal{M}_2$, $\mathcal{M}_3$}, 
    error bars/y dir=both,
    error bars/y explicit,
]

\addplot+[
    error bars/.cd,
    y dir=both,
    y explicit,
] coordinates {
    (4, -36.16) +- (8.13, 8.13)
    (8, -31.07) +- (16.12, 16.12)
    (16, -58.46) +- (5.05, 5.05)
    (32, -79.49) +- (36.19, 36.19)
    (64, -394.11) +- (28.14, 28.14)
    (128, -430.77) +- (330.48, 330.48)
    (256, -812.91) +- (188.22, 188.22)
    (512, -779.42) +- (298.02, 298.02)
    (1024, -876.48) +- (313.52, 313.52)
};

\addplot+[
    error bars/.cd,
    y dir=both,
    y explicit
] coordinates {
    (4, -44.59) +- (13.59, 13.59)
    (8, -54.27) +- (31.70, 31.70)
    (16, -55.78) +- (2.17, 2.17)
    (32, -98.22) +- (3.83, 3.83)
    (64, -35.65) +- (16.10, 16.10)
    (128, -59.35) +- (16.82, 16.82)
    (256, -120.82) +- (86.10, 86.10)
    (512, -116.78) +- (87.55, 87.55)
    (1024, -238.08) +- (60.05, 60.05)
}; 

\addplot+[
    error bars/.cd,
    y dir=both,
    y explicit,
] coordinates {
    (4, -34.16) +- (10.41, 10.41)
    (8, -33.07) +- (19.32, 19.32)
    (16, -52.54) +- (2.05, 2.05)
    (32, -54.35) +- (2.12, 2.12)
    (64, -28.78) +- (15.67, 15.67)
    (128, -47.72) +- (15.52, 15.52)
    (256, -132.29) +- (72.65, 72.65)
    (512, -117.67) +- (76.20, 76.20)
    (1024, -178.32) +- (84.21, 84.21)
};

\end{axis}
\end{tikzpicture}
    }
    \caption{Observed cumulative reward scores during the evaluation of the trained models on different number of topologies. The 4 - 128 topologies are evaluated for 1000 episodes while the 256 - 1024 topologies are evaluated for 50 episodes. Each episode is comprised of 100 steps. The tabular version of this data can be found in the supplement, \cref{tab:gacd_topologies}.}
    \label{fig:number_of_topologies_vs_reward}
\end{figure}

We evaluate $\{\mathcal{M}_i\}_{i=1}^3$ against a varying number of network topologies, ranging from 4 to 1024, randomly sampled from the \psg. For model $\mathcal{M}_{3}$, we sample 10 episodes of 50 steps for each other network topology to create the different graph representations of the network to pre-train the \grl agent. Each model is trained on these environments against a randomly sampled red agent, choosing from either the B-Line or the Meander agent from the \cage environment. The performance of the \gacd agents are illustrated in \cref{fig:number_of_topologies_vs_reward}. We observe that the performance of the \gacd agent degrades with an increasing number of topologies. Next, the use of a \gtm improves the performance compared to $\mathcal{M}_1$, with the use of \gtm in the auto-encoding setup performing the best. $\mathcal{M}_1$ shows poor scaling with the number of topologies, compared with $\mathcal{M}_{2}$ and $\mathcal{M}_{3}$. $\mathcal{M}_{2}$ and $\mathcal{M}_{3}$ that use the \gtm architecture achieve performance comparable to the \sota \ppo model, from Cardiff University \citep{vyas2023automated}, for up to 128 topologies, while the $\mathcal{M}_{1}$ only shows performance comparable to the \sota on 16 topologies.

\subsection{Ablation study}
\label{sec:ablation_study}
We perform an ablation study by where the different components of \gacd are analyzed in isolation where we use $\mathcal{M}_1$ and $\mathcal{M}_2$ as reference. Results are shown in \cref{tab:model_ablation} and it can be see that in general it is better to add more layers to each layer type. An expected finding.
\begin{table}[ht!]
\centering
\caption{Ablation study comparing different graph neural network architectures and their depths. Each setting was trained on 32 topologies and evaluating on 50 randomly sampled ones.}
\label{tab:model_ablation}
    \resizebox{\columnwidth}{!}{ 
    \begin{tabular}{cccc}
    \toprule
    Model variant & Layer type & \#Layers & Reward \\
    \midrule
    $\mathcal{M}_1$ & GCN & 1 & $-182.04 \pm 89.71$ \\
    & & 2 & $-155.11 \pm 44.56$ \\
    & GAT & 1 & $-214.71 \pm 56.55$ \\
    & & 2 & $-61.62 \pm 16.87$ \\
    & GCN + Graph Embed. & 1 & $-192.53 \pm 59.29$ \\
    & & 2 & $-98.81 \pm 35.93$ \\
    $\mathcal{M}_2$ & Graphormer & 1 & $-31.82 \pm 30.95$ \\
    & & 2 & $-18.35 \pm 9.78$ \\
    \bottomrule
    \end{tabular}
    }
\end{table}

\subsection{Scaling to larger networks}
\label{sec:scaling_to_larger_networks}
In this experiment we scale the \cage network to have 1000 hosts across 10 subnets, with 100 hosts per subnet, across 50 episodes per red agent. $\mathcal{M}_3$ is used for evaluation, results are shown in \cref{tab:red_agent_comparison}.
\begin{table}[hb!]
\centering
\caption{Results for $\mathcal{M}_3$ when \cage has been scaled significantly in size to have 1000 hosts across 10 different subnets. Evaluation is shown against two types of red-agent.}
\label{tab:red_agent_comparison}
    \begin{tabular}{cc}
    \toprule
    Red agent & Reward \\
    \midrule
    Meander & $-2.40 \pm 5.65$ \\
    B-Line & $-50.77 \pm 20.45$ \\
    \bottomrule
    \end{tabular}
\end{table}

As is evident in the Meander case, designing a good red agent on large networks is a difficult task. Even in the B-Line red agent setting, we only observe 23 impacts, i.e., a successful infection of the operational server, across 50 episodes, each of 100 steps.

\subsection{Novel attack patterns}
\label{sec:novel_pattern}

Next we explore novel attack patterns. We conduct two experiments where in the first, we retrain the \gacd ($\mathcal{M}_2$) against only the B-Line agent and evaluate against the Meander red agent, and in the second experiment we train the \gacd agent against the Meander red agent and evaluate against B-Line. \Cref{tab:red_agent_cross_evaluation} documents the observed performance.

\begin{table}[ht!]
\centering
\caption{Cross-evaluation performance between different red agent strategies.}
\label{tab:red_agent_cross_evaluation}
    \resizebox{\columnwidth}{!}{
    \begin{tabular}{ccc}
    \toprule
    Trained red agent & Evaluation red agent & Reward \\
    \midrule
    Meander & B-Line & $-15.93 \pm 5.24$ \\
    B-Line & Meander & $-20.76 \pm 14.52$ \\
    \bottomrule
    \end{tabular}
    }
\end{table}

As noted in \cref{sec:scaling_to_larger_networks} where we scale our method to larger networks; designing a red agent is a challenging task. Hence, we limit our analysis to evaluations on the red agents provided as part of \cage.

\subsection{Comparison against other \gnn models}
\label{sec:comparison_sota}

The $ \mathcal{M}_3 $ variant of \gacd is evaluated against the \sota \gnn-based cyber defense agent from \citet{nyberg2024structural} on the seven environments from \citet{nyberg2024structural}, where subnet sizes vary from 10 to 16 hosts while maintaining the same network topology. We compare \gacd against the \mpnn-2 model from \citet{nyberg2024structural}, trained for $ 0.8 \times 10^6 $ time steps, evaluated against the Meander red agent. In contrast, \gacd’s $ \mathcal{M}_3 $ encoder is pre-trained on 256 randomly sampled topologies from the \psg and fine-tuned on the 7 test environments, training for $2 \times 10^6 $ timesteps with random red agent selection. Results in \cref{tab:ex3_average_cumulative_reward} show that, on average, \gacd outperforms the the \mpnn agents from \citet{nyberg2024structural}, on all subnet sizes.
\begin{table*}[ht!]
    \centering
    \renewcommand{\arraystretch}{1.2} 
    \setlength{\tabcolsep}{6pt} 
    \caption{Average cumulative reward (100 steps) within one standard deviation, measured over 1000 episodes. The score for the top ranking agent for a given number of hosts is shown in bold and the score for the agent ranking second is italicized.}
    \label{tab:ex3_average_cumulative_reward}
    \resizebox{\textwidth}{!}{ 
    \begin{tabular}{cccccccccc}
    \toprule
    \# Hosts & \gacd $(\mathcal{M}_3)$ & \mpnn G-2-10 & \mpnn G-2-11 & \mpnn G-2-12 & \mpnn G-2-13 & \mpnn G-2-14 & \mpnn G-2-15 & \mpnn G-2-16 \\
    \midrule
        $10$ & $\mathbf{-15.00} \pm \mathbf{3.67}$ & $-\textit{28.80} \pm \textit{7.09}$ & $-89.04 \pm 28.27$ & $-74.23 \pm 40.05$& $-75.32 \pm 19.27$ & $-92.68 \pm 47.57$ & $-69.44 \pm 6.02$ & $-64.97 \pm 1.71$ \\
        $11$ & $\mathbf{-15.64} \pm \mathbf{3.23}$ & $-57.03 \pm 19.45$ & $-\textit{36.01} \pm \textit{17.74}$ & $-70.99\pm 48.01$& $-92.27 \pm 26.63$ & $-92.59 \pm 50.19$ & $-61.64 \pm 10.73$ & $-64.6 \pm 3.25$ \\
        $12$ & $\mathbf{-15.77} \pm \mathbf{3.28}$ & $-50.32 \pm 15.57$ & $-94.41 \pm 41.39$ & $-\textit{25.26} \pm \textit{19.78}$ & $-78.39 \pm 21.55$ & $-76.72 \pm 36.51$ & $-65.14 \pm 9.98$ & $-57.99 \pm 2.55$ \\
        $13$ & $\mathbf{-19.39} \pm \mathbf{4.04}$ & $-69.97 \pm 21.64$ & $-85.71 \pm 26.25$ & $-72.42\pm 41.45$& $-\textit{32.67} \pm \textit{13.47}$ & $-82.25 \pm 42.65$ & $-63.91 \pm 10.27$ & $-69.23 \pm 0.73$ \\
        $14$ & $\mathbf{-17.41} \pm \mathbf{3.30}$ & $-71.27 \pm 12.88$ & $-79.22 \pm 24.14$ & $-67.69\pm 33.06$& $-84.57 \pm 18.40$ & $-\textit{24.61} \pm \textit{14.95}$ & $-67.9 \pm 4.09$ & $-70.83 \pm 0.94$ \\
        $15$ & $\mathbf{-17.38} \pm \mathbf{3.18}$ & $-49.09 \pm 7.72$ & $-65.75 \pm 21.81$ & $-57.85\pm 30.33$& $-70.32 \pm 9.92$ & $-73.23 \pm 43.26$ & $-\textit{22.59} \pm \textit{5.99}$ & $-56.67 \pm 2.08$ \\
        $16$ & $\mathbf{-16.39} \pm \mathbf{3.25}$ & $-56.98 \pm 19.52$ & $-82.41 \pm 33.22$ & $-50.70\pm 30.03$& $-72.47 \pm 15.54$ & $-69.92 \pm 36.24$ & $-67.29 \pm 17.93$ & $-\textit{18.81} \pm \textit{5.59}$ \\
        \midrule
        Mean & $\mathbf{-16.71} \pm \mathbf{3.42}$ & $-54.78 \pm 14.83$ & $-76.07 \pm 27.54$ & $-59.88\pm 24.13$& $-72.28 \pm 17.82$ & $-73.14 \pm 38.77$ & $-57.41 \pm 10.10$ & $-\textit{57.26} \pm \textit{3.17}$ \\
    \bottomrule
    \end{tabular}
    }
\end{table*}

\subsection{Comparing information representations}
\label{sec:randomization}

To demonstrate the importance of a graph-encoded representation of the network, we compare the performance of the \gacd agent against the \sota \ppo algorithm \citep{vyas2023automated} that uses a flattened representation of the observation. We also include results for the various \mpnn variations to explore how they far in this experimental scenario.

For this comparison, we compare the performance of the \gacd agent on vanilla \cage scenario, involving 13 hosts, in a setting that introduces a randomization to the sequence of the nodes in the network. Here both the \sota \ppo and \gacd agents are trained for $2 \times 10^6$ time-steps, with a single \gacd agent being trained against both the B-Line and Meander red agents and two different agents are trained employing the \sota \ppo algorithm. The results comparing the performance in this setting are presented in \cref{tab:gacd_vs_cardiff_randomized} where we observe that the $\mathcal{M}_{3}$ variant of \gacd is able to successfully adapt to the randomization in the node sequence, while the \sota \ppo agent fails to adapt to such modification.
\begin{table}[ht!]
\centering
\caption{Average cumulative reward (100 Steps) within one standard deviation. The score for the best performing model is shown in bold. Both the Cardiff \ppo and the \gacd agents are trained without node sequence randomization and evaluated with randomized sequence of nodes.}
\label{tab:gacd_vs_cardiff_randomized}
    \resizebox{\columnwidth}{!}{ 
    \begin{tabular}{llll}
      \toprule
      Blue agent & Red agent & W/o randomization & With randomization \\
      \midrule
          \gacd $(\mathcal{M}_3)$ & B-Line & $-41.78 \pm 20.17$ & $-41.78 \pm 20.17$ \\
          \gacd $(\mathcal{M}_3)$ & Meander & $-20.49 \pm 10.48$ & $ \mathbf{-20.49} \pm \mathbf{10.48} $ \\
          Cardiff-\ppo & B-Line & $\mathbf{-13.23} \pm \mathbf{4.24}$ & $-699.13 \pm 425.53$ \\
          Cardiff-\ppo & Meander & $-16.28 \pm 3.82$ & $-520.54 \pm 312.06$ \\
            \mpnn G-2-10 & B-Line & $-68.32 \pm 38.29$ & $-68.32 \pm 38.29$ \\
            \mpnn G-2-10 & Meander & $-28.98 \pm 8.87$ & $-28.98 \pm 8.87$ \\
            \mpnn G-2-11 & B-Line & $-93.72 \pm 28.79$ & $-93.72 \pm 28.79$ \\
            \mpnn G-2-11 & Meander & $-33.56 \pm 9.06$ & $-33.56 \pm 9.06$ \\
            \mpnn G-2-12 & B-Line & $-23.43 \pm 91.48$ & $-23.43 \pm 91.48$ \\
            \mpnn G-2-12 & Meander & $-32.07 \pm 15.78$ & $-32.07 \pm 15.78$ \\
            \mpnn G-2-13 & B-Line & $-99.20 \pm 19.16$ & $-99.20 \pm 19.16$ \\
            \mpnn G-2-13 & Meander & $-31.70 \pm 7.83$ & $-31.70 \pm 7.83$ \\
            \mpnn G-2-14 & B-Line & $-66.15 \pm 5.58$ & $-66.15 \pm 5.58$ \\
            \mpnn G-2-14 & Meander & $-31.70 \pm 7.83$ & $-31.70 \pm 7.83$ \\
            \mpnn G-2-15 & B-Line & $-67.35 \pm 11.07$ & $-67.35 \pm 11.07$ \\
            \mpnn G-2-15 & Meander & $-31.34 \pm 6.25$ & $-31.34 \pm 6.25$ \\
            \mpnn G-2-16 & B-Line & $-88.36 \pm 28.03$ & $-88.36 \pm 28.03$ \\
            \mpnn G-2-16 & Meander & $-31.47 \pm 6.93$ & $-31.47 \pm 6.93$ \\
      \bottomrule
    \end{tabular}
    }
\end{table}

\subsection{Model adaptability to environment dynamics}
\label{sec:dynamic_env}

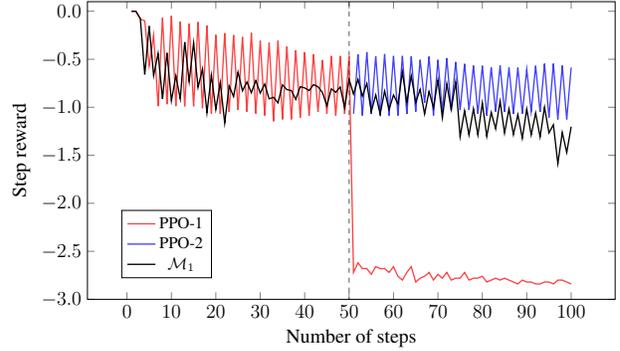
\begin{figure}[ht!]
    \centering
    \resizebox{\columnwidth}{!}{
        \begin{tikzpicture}
\begin{axis}[
    width=0.75\textwidth,
    height=0.75\textwidth / 1.618,
    line width=0.5,
    grid=major, 
    tick label style={font=\large},
    legend style={nodes={scale=1, transform shape}},
    label style={font=\large},
    legend image post style={},
    grid style={white},
    xlabel={Number of steps},
    ylabel={Step reward},
    y tick label style={
      /pgf/number format/.cd,
      fixed,
      fixed zerofill,
      precision=1
   },
  legend style={at={(0.15, 0.3)}, anchor=north, fill=none},
  ymax = 0.1,
  ymin = -3.,
    ]

    \addplot[red, thick,opacity=0.75] coordinates {
        (1, 0.0) (2, 0.0) (3, -0.082) (4, -0.1) (5, -0.568)
        (6, -0.248) (7, -0.988) (8, -0.068) (9, -0.968) (10, -0.048)
        (11, -0.988) (12, -0.108) (13, -0.99) (14, -0.13) (15, -0.99)
        (16, -0.112) (17, -1.01) (18, -0.15) (19, -1.066) (20, -0.246)
        (21, -1.068) (22, -0.228) (23, -1.048) (24, -0.248) (25, -1.028)
        (26, -0.268) (27, -1.046) (28, -0.226) (29, -1.066) (30, -0.306)
        (31, -1.086) (32, -0.328) (33, -1.146) (34, -0.306) (35, -1.126)
        (36, -0.366) (37, -1.126) (38, -0.406) (39, -1.106) (40, -0.426)
        (41, -1.086) (42, -0.448) (43, -1.048) (44, -0.508) (45, -1.066)
        (46, -0.466) (47, -1.086) (48, -0.466) (49, -1.068) (50, -0.468)
        (51, -2.72) (52, -2.62) (53, -2.68) (54, -2.68) (55, -2.74)
        (56, -2.66) (57, -2.68) (58, -2.68) (59, -2.72) (60, -2.66)
        (61, -2.76) (62, -2.8) (63, -2.72) (64, -2.66) (65, -2.82)
        (66, -2.78) (67, -2.76) (68, -2.72) (69, -2.8) (70, -2.74)
        (71, -2.78) (72, -2.72) (73, -2.8) (74, -2.78) (75, -2.78)
        (76, -2.72) (77, -2.8) (78, -2.78) (79, -2.78) (80, -2.76)
        (81, -2.82) (82, -2.8) (83, -2.78) (84, -2.8) (85, -2.78)
        (86, -2.8) (87, -2.82) (88, -2.84) (89, -2.8) (90, -2.82)
        (91, -2.82) (92, -2.84) (93, -2.84) (94, -2.82) (95, -2.82)
        (96, -2.84) (97, -2.8) (98, -2.8) (99, -2.82) (100, -2.84)
    };
    \addlegendentry{PPO-1}

    \addplot[blue, thick,opacity=0.75] coordinates {
        (51, -1.068) (52, -0.448) (53, -1.086) (54, -0.426) (55, -1.068)
        (56, -0.468) (57, -1.086) (58, -0.446) (59, -1.086) (60, -0.466)
        (61, -1.086) (62, -0.466) (63, -1.07) (64, -0.47) (65, -1.046)
        (66, -0.506) (67, -0.952) (68, -0.472) (69, -1.086) (70, -0.506)
        (71, -1.046) (72, -0.488) (73, -1.028) (74, -0.528) (75, -1.028)
        (76, -0.548) (77, -1.028) (78, -0.566) (79, -1.068) (80, -0.568)
        (81, -1.086) (82, -0.586) (83, -1.046) (84, -0.566) (85, -1.046)
        (86, -0.586) (87, -1.068) (88, -0.588) (89, -1.066) (90, -0.606)
        (91, -1.046) (92, -0.568) (93, -1.048) (94, -0.546) (95, -1.106)
        (96, -0.546) (97, -1.126) (98, -0.566) (99, -1.126) (100, -0.586)
    };
    \addlegendentry{PPO-2}

    \addplot[black, thick] coordinates {
        (1, 0.0) (2, 0.0) (3, -0.076) (4, -0.658) (5, -0.154)
        (6, -0.67) (7, -0.44) (8, -0.912) (9, -0.434) (10, -0.932)
        (11, -0.672) (12, -0.912) (13, -0.322) (14, -0.732) (15, -0.33)
        (16, -0.792) (17, -0.602) (18, -1.056) (19, -0.666) (20, -1.022)
        (21, -0.744) (22, -1.164) (23, -0.778) (24, -0.888) (25, -0.634)
        (26, -0.828) (27, -0.736) (28, -0.848) (29, -0.722) (30, -0.852)
        (31, -0.758) (32, -0.912) (33, -0.896) (34, -0.954) (35, -0.766)
        (36, -0.816) (37, -0.826) (38, -0.918) (39, -0.786) (40, -0.798)
        (41, -0.826) (42, -0.764) (43, -0.786) (44, -0.846) (45, -0.984)
        (46, -0.808) (47, -0.844) (48, -0.788) (49, -1.004) (50, -0.708)
        (51, -0.862) (52, -0.71) (53, -0.884) (54, -0.752) (55, -1.044)
        (56, -0.872) (57, -1.07) (58, -0.832) (59, -1.024) (60, -0.872)
        (61, -0.966) (62, -0.632) (63, -0.948) (64, -0.674) (65, -0.948)
        (66, -0.814) (67, -1.07) (68, -0.836) (69, -1.004) (70, -0.716)
        (71, -1.108) (72, -0.756) (73, -1.022) (74, -0.736) (75, -1.306)
        (76, -1.036) (77, -1.288) (78, -1.016) (79, -1.288) (80, -0.956)
        (81, -1.302) (82, -1.054) (83, -1.224) (84, -0.938) (85, -1.268)
        (86, -1.018) (87, -1.284) (88, -1.076) (89, -1.322) (90, -1.02)
        (91, -1.288) (92, -1.06) (93, -1.308) (94, -1.0) (95, -1.266)
        (96, -1.182) (97, -1.586) (98, -1.262) (99, -1.47) (100, -1.202)
    };
    \addlegendentry{$\mathcal{M}_1$}

    \addplot[dashed, thick, opacity=0.5] coordinates {(50, 0.1) (50, -3)};
\end{axis}
\end{tikzpicture}
    }
    \caption{Comparison of the averaged step-wise reward returned from a discontinuous environment for the \gacd agent vs the baseline \ppo agent.}
    \label{fig:trained_agent_adaptation}
\end{figure}

We evaluate the inference capability of the $ \mathcal{M}_1 $ \gacd agent by introducing a network topology change midway through an episode. The agent is trained on 8 topologies for $ 2 \times 10^6 $ timesteps, and during evaluation, the network shifts from the vanilla \cage topology to one of the trained topologies. Adaptability is assessed over 1000 episodes of 100 steps, with average rewards shown in \cref{fig:trained_agent_adaptation}. Compared to the \sota \ppo agent, which becomes ineffective post-change (red curve, receiving the minimum reward of $-3$), the \gacd agent (black curve) adapts seamlessly with minimal reward degradation. We also evaluate a hybrid approach where a secondary \ppo agent is pre-trained on the new topology (blue curve). If the defense policy switches at the right moment, the combined \ppo agents perform comparably to a single \gacd agent.

The results are available in tabular form in \cref{tab:step_reward_comparison}, where we have also included comparison against one of the \mpnn variations.
\begin{table}[ht!]
\centering
\caption{Average step-wise reward comparison across different models at select step of the episodes.}
\label{tab:step_reward_comparison}
\resizebox{\columnwidth}{!}{
    \begin{tabular}{ccccc}
    \toprule
    Step size & $\mathcal{M}_1$ & \mpnn G-2-13 & \ppo-1 & \ppo-2 \\
    \midrule
    25 & $-0.634$ & $-0.813$ & $-1.028$ & N/A \\
    50 & $-0.708$ & $-0.921$ & $-0.468$ & N/A \\
    75 & $-1.306$ & $-0.883$ & $-2.780$ & $-1.028$ \\
    100 & $-1.202$ & $-1.065$ & $-2.840$ & $-0.586$ \\
    \bottomrule
    \end{tabular}
}
\end{table}

\subsection{Optimal transport utility}
\label{sec:optimal_transport_utility}

Our \gacd model without the \ot layer reduces to MPNN i.e. a standard \gnn-based model. We provide a comparison of the performance of $\mathcal{M}_3$ and MPNN-G-2-13, evaluated in the 64 topology setting against both red agents, randomly sampled, for 100 episodes each, in \cref{tab:ot_utility_table}.
\begin{table}[ht!]
\centering
\caption{Measured results when the optimal transport layer has been turned off for the $\mathcal{M}_3$ model. For comparison, the \mpnn-G-2-13 model is also included in the table.}
\label{tab:ot_utility_table}
    \begin{tabular}{ll}
    \toprule
    Blue agent & Reward \\
    \midrule
    \gacd $(\mathcal{M}_3)$ & $-28.78 \pm 15.67$ \\
    \mpnn-G-2-13 & $-69.20 \pm 16.81$ \\
    \bottomrule
    \end{tabular}
\end{table}

The primary advantage of $\mathcal{M}_3$ arises from the fact that the embedding model is trained offline from the \rrl agent. As a result, the agent first learns to structure the latent space of topologies before learning to take the right action for each topology. In the case of $\mathcal{M}_2$, where the \ot-based embedding layers and the \rrl action network are trained together, the agent prioritizes the \ppo loss which can lead to imperfect latent spaces, which $\mathcal{M}_3$ is not affected by.

\section{Conclusion and discussion}
\label{sec:conclusion}


Comparisons with existing methods show that \gacd performs on par with other graph-based agents that do not utilize optimal transport, while offering greater adaptability to varying network topologies and adversarial strategies. Unlike standard approaches that encode rich state information as flattened observations, our graph-based representation enables a more generalizable policy that dynamically adapts to environmental changes. However, \gacd incurs a higher training cost, requiring 3–10× more training to match the performance of other \sota agents. Additionally, we found that reward shaping across different environments is critical for optimal performance. Moving forward, incorporating domain knowledge into \acd agents could improve explainability, efficiency and robustness, guiding decision-making with additional structured models.


\section{Acknowledgments}

This research is supported by the Defense Advanced Research Project Agency (\textsc{darpa}) through the `Cyber Agents for Security Testing and Learning Environments' (\textsc{castle}) program under Contract No. \texttt{W912CG23C0029}. The views, opinions and/or findings expressed are those of the author and should not be interpreted as representing the official views or policies of the Department of Defense or the U.S. Government.
\bibliography{content/paper}
\bibliographystyle{icml2025}

\newpage
\appendix
\onecolumn
\section{Experimental data}
\begin{table}[ht!]
\centering
\caption{Average cumulative reward (100 Steps) within one standard deviation. The score for the best performing model is shown in bold for that number of topologies. Results are also shown graphically in \cref{fig:number_of_topologies_vs_reward}.}
\label{tab:gacd_topologies}
    \begin{tabular}{cccc}
      \toprule
      \# Topologies & $\mathcal{M}_1$ & $\mathcal{M}_2$ & $\mathcal{M}_3$ \\
      \midrule
          $4$ & $-36.16 \pm 8.13$ & $-44.59 \pm 13.59$ & $\mathbf{-34.16} \pm \mathbf{10.41}$ \\
          $8$ & $\mathbf{-31.07} \pm \mathbf{16.12}$ & $-54.27 \pm 31.70$ & $-33.07 \pm 19.32$ \\
          $16$ & $-58.46 \pm 5.05$ & $-55.78 \pm 2.17$ & $\mathbf{-52.54} \pm \mathbf{2.05}$ \\
          $32$ & $-79.49 \pm 36.19$ & $-98.22 \pm 3.83$ & $\mathbf{-54.35} \pm \mathbf{2.12}$ \\
          $64$ & $-394.11 \pm 28.14$ & $-35.65 \pm 16.10$ & $\mathbf{-28.78} \pm \mathbf{15.67}$ \\
          $128$ & $-430.77 \pm 330.48$ & $-59.35 \pm 16.82$ & $\mathbf{-47.72} \pm \mathbf{15.52}$ \\
          $256$ & $-812.91 \pm 188.22$ & $\mathbf{-120.82} \pm \mathbf{86.10}$ & $-132.29 \pm 72.65$ \\
          $512$ & $-779.42 \pm 298.02$ & $\mathbf{-116.78} \pm \mathbf{87.55}$ & $-117.67 \pm 76.20$ \\
          $1024$ & $-876.48 \pm 313.52$ & $-238.08 \pm 60.05$ & $\mathbf{-178.32} \pm \mathbf{84.21}$ \\
      \bottomrule
    \end{tabular}
\end{table}


\section{Fused Gromov-Wasserstein distance}
\label{sec:fgw_distance}

\citet{vayer2018optimal} suggest a paradigm for viewing graphs as probability distributions, embedded in a specific metric space. Notably, in their work they operate on attributed graphs which is to say structured data with both feature information as well as structure information. Optimal transport (\ot) provides a principled framework for comparing probability measures $\mu$  and  $\nu$  by determining the most efficient way to transform one into the other \citep{thorpe2018introduction}. Specifically, it seeks an optimal coupling --also known as a transport plan-- that minimizes the total transport cost required to move mass from  $\mu$  to  $\nu$. The resulting minimal cost defines the optimal transport distance, which serves as a meaningful metric for comparing distributions \citep{wang2024galopa}. 

We are interested in the distance between two graphs $G_1$ and $G_2$ with associated feature sets $\X_1$ and $\X_2$ respectively, each described by their probability measure, as noted above:
\begin{align}
    \mu = \sum_{i=1}^n h_i \delta_{(x_i, a_i)} \\
    \nu = \sum_{i=1}^n g_j \delta_{(y_j, a_j)}
\end{align}
where $h_i \in \sum_n$ and $g_j \in \sum_m$ are histograms, $\delta_x$ is the Dirac delta centered on $x$. Let $\Pi(h,g)$ be the set of all admissible couplings (a way to describe a joint probability distribution over two spaces that respects given marginal distributions) between $h$ and $g$, in other words the set:
\begin{equation}
    \Pi(h,g) = \{\pi \in \mathbb{R}_+^{n\times m} s.t. \sum_{i=1}^n = \pi_{i,j} = h_j; \sum_{j=1}^m = \pi_{i,j} = g_i\}
\end{equation}
where $\pi_{i,j}$ is the amount of mass shifted from bin $h_i$ to $g_j$ for coupling\footnote{Intuition: given $\mu$  and  $\nu$, a coupling tells you how to pair samples from $\mu$  with samples from $\nu$.} $\pi$ \citep[\S 3]{vayer2018optimal}. Where we can construe $\pi$ as a matrix which describes the probabilistic matching of the nodes of $G_1$ and $G_2$. Further, let $M_{AB} = (d(a_i,b_j))_{i.j}$ be a $n\times m$ matrix which measures the distance between the features according to some distance $d(\cdot,\cdot)$. Then the structure matrices\footnote{\citet{vayer2018optimal} abuses notation somewhat, $C$ is a function that measures the similarity of node attributes, $C$ is also the structure matrix which stores these similarities.} are denoted $C_1 = C(\x_i,\x_j), \forall \x_i,\x_j \in \X_1$ and  $C_2 = C(\x_i,\x_j), \forall \x_i,\x_j \in \X_2$. Where $\mu_X$ and $\mu_A$ as well as we $\nu_Y$ and $\nu_B$ represent the marginals of $\mu$ and $\nu$ respectively. Finally \citet{vayer2018optimal} define a 4-dimensional tensor which measures the similarity between two structure matrices i.e.
\begin{equation}
    L_{i,j,k,l} = \| C_1(i,k) - C_2(j,l) \|.
\end{equation}
We now have all the ingredients we need to define the Fused Gromov-Wasserstein distance which is defined given a trade-off parameter $\alpha \in [0,1]$ and where the metric is defined for $q=1$ (but is defined also for larger values but that discussion is outside the scope of this review):
\begin{equation}
    d_{\fgw}^{\alpha,q} = \min_{\pi \in \Pi(h,g)} E_q (M_{AB},C_1,C_2, \pi)
\end{equation}
where
\begin{align}
    E_q (M_{AB},C_1,C_2, \pi) &= \langle (1-\alpha) M^q_{AB} + \alpha L (C_1,C_2)^q \otimes \pi, \pi \rangle \\
    &= \sum_{i,j,k,l} (1-\alpha) d(a_i,b_j)^q + \alpha \| C_1(i,k) - C_2(j,l)\|^q \pi_{i,j} \pi_{k,l}.
\end{align}
The \fgw distance jointly exploits both features and structure.

\section{Procedural scenario generator algorithm}
\label{sec:psg}

To create structural variations in the \cage environment network topology, we employ a procedural scenario generator (\psg) that randomly synthesized new environments following a set of rules with which to guide the environment dynamics. \cref{alg:procedural_scenario_generator} presents the steps involved in the \psg process, where we randomly sample the number of subnets, the number of hosts on each subnet and the assignment of the agents to subnets hosts based on an input of the upper and lower bounds for these. The generator creates and returns the graph representation of the network based on the configurations provided.


\begin{algorithm}[ht!]
    \caption{Procedural scenario generator}\label{alg:procedural_scenario_generator}
    \begin{algorithmic}[1]
        \STATE \textbf{function} \textbf{CreateAgent}($H_i$):
        \STATE Agent: $a = \mathbf{newAgent}(H_i)$
        \STATE \textbf{return} $a$
        \STATE \textbf{end function}
        \item[]
        \STATE \textbf{function} \textbf{CreateSubnet}($ST$, $N_H$):
        \STATE $\mathcal{S} = \mathbf{newSubnet}[ST]$
        \FOR{\texttt{$n = 0, \ldots, N_H$}}
            \STATE ${HT}_i = \mathbf{C}(\{HT\}, 1)$
            \STATE $H_n = \mathbf{CreateHost}({HT}_i)$
            \STATE $S \to H := S \to H \space \bigcup \space \{H_n\}$
        \ENDFOR
        \STATE \textbf{return} $\mathcal{S}$
        \STATE \textbf{end function}
        \item[]
        \STATE \textbf{function} \textbf{CreateHost}($HT$):
        \STATE $H = \mathbf{newHost}[HT]$
        \STATE \textbf{return} $H$
        \STATE \textbf{end function}
        \item[]
        \STATE \textbf{Input:} Definition of minimum and maximum number of subnets (${NS}_l, {NS}_u$), the minimum and maximum number of hosts in the network (${NH}_l, {NH}_u$), a default set of subnets ($\{S\}$) the actions associated with the red, blue and green agents ($a_R, a_B, a_G$)
        \STATE \textbf{Output:} The generated network topology $\mathcal{G}$ – comprised of a set of subnets ($\{S\}$), a set of hosts ($\{H\}$), and a set of agents ($\{A\}$).
        \item[]
        \STATE Number of subnets: $NS \sim \mathcal{U}[{NS}_l, {NS}_u]$
        \STATE Number of hosts: $NH \sim \mathcal{U}[\max(NS, {NH}_l), {NH}_u]$
        \STATE Subnet Options: $O = \{User, Operational, Enterprise\}$
        \FOR{$i = 0, \ldots, NS$}
            \STATE Subnet type: ${ST}_i = \mathbf{C}(O, 1)$
        \ENDFOR
        \FOR{$i = 0, \ldots, NS$}
            \STATE $\mathcal{S}_i = \mathbf{CreateSubnet}(ST_i, NH_i)$
        \ENDFOR
        \STATE Enterprise subnet: $\mathcal{S}_E = \mathbf{C}(\{\mathcal{S}_1, \ldots, \mathcal{S}_{NS}\}, 1)$
        \STATE Operational subnet: $\mathcal{S}_O = \mathbf{C}(\{\mathcal{S}_1, \ldots, \mathcal{S}_{NS}\} \space \backslash \space \mathcal{S}_E, 1)$
        \item[]
        \STATE Red Agent: $a_R = \mathbf{CreateAgent(\mathcal{S}_0 \to \mathcal{H}_0)}$
        \STATE Blue Agent: $a_B = \mathbf{CreateAgent(\mathcal{S}_E \to \mathcal{H}_0)}$
        \STATE Green Agent: $a_G = \mathbf{CreateAgent(C(\{H\}, 1))}$
        \STATE \textbf{return} The generated network topology $\mathcal{G} := \{\{S\}, \{A\}, \{H\}\}$, where $\{A\} = \{a_R, a_B, a_G\}$
    \end{algorithmic}
\end{algorithm}
\clearpage

\section{Understanding the distributions in the commutative diagram}
\label{sec:map_analogy}

Imagine you want to draw a map of all the cities in the USA.
\begin{itemize}
    \item[$\mu_{\gt}$] This would be the true, ideal probability distribution of \textit{all actual human settlements} across the USA (cities, towns, villages, etc.), which is complex and includes lots of clusters and empty spaces. You can never perfectly know this.
    \item[$\hat{\mu}_{\gt}$] This is the \textit{actual data} -- the specific locations of the 100 largest cities you have in your dataset. This is a finite and discrete list of points.
    \item[$\mu$] This is the learned ``map'' of settlement density. We take a simple grid (the initial noise $\mu$ for the \sdot) and warp it such that the density on the warped map corresponds to the locations of the 100 cities. This warped map $(\mu)$ now has high-density areas where cities are and low-density areas in between, accurately reflecting the real distribution, even though it is still a continuous ``map''. Now, if we give this smart, warped map to someone to `fill in the details' (our neural-net model), they will do a much better job than if we just gave them a blank, uniform grid.
\end{itemize}


\end{document}